\documentclass[journal=jctcce,manuscript=article]{achemso}
\setkeys{acs}{articletitle=true}
\usepackage{graphicx}
\usepackage{amsmath}
\usepackage{amsfonts}
\usepackage{amssymb}
\usepackage{amsthm}
\usepackage{braket}
\usepackage{multirow}
\usepackage{achemso}
\usepackage[labelformat=empty]{subfig}
\usepackage{bm}%lettere greche in grassetto, e.g. $\bm{\rho}$
\usepackage{booktabs}
\usepackage{comment}
\usepackage{upgreek}

\usepackage[utf8]{inputenc}
\newlength\tindent
\newcommand{\parz}[2]{ \frac{\partial{#1}}{\partial{#2}}}            %derivate parziali
\newcommand{\abs}[1]{\left| {#1} \right|}              % abb. sbarrette valore assoluto

\newcommand{\gr}[1]{\textbf{#1}}

\newcommand{\bq}{\mathbf{q}}
\newcommand{\bS}{\mathbf{S}}

\newcommand{\bV}{\mathbf{V}}

\newcommand{\bD}{\mathbf{D}}

\newcommand{\bR}{\mathbf{R}}

\newcommand{\bP}{\mathbf{P}}

\newcommand{\bF}{\mathbf{F}}

\newcommand{\bW}{\mathbf{W}}

\newcommand{\Pmn}{P_{\mu\nu}}
\newcommand{\Gmn}{G_{\mu\nu}}

\newcommand{\Fmnt}{\widetilde{F}_{\mu\nu}}
\newcommand{\hmn}{h_{\mu\nu}}

\newcommand{\bVmn}{\mathbf{V}_{\mu\nu}}

\newcommand{\bmu}{\boldsymbol{\mu}}

\newcommand{\bL}{\mathbf{L}}
\newcommand{\bE}{\mathbf{E}}
\newcommand{\bEmn}{\mathbf{E}_{\mu\nu}}
\newcommand{\bRmn}{\mathbf{R}_{\mu\nu}}

\DeclareMathOperator\erf{erf}

\author{Tommaso Giovannini}
\affiliation{Department of Chemistry, Norwegian University of Science and Technology, 7491 Trondheim, Norway}              
\author{Laura Grazioli}
\affiliation{Scuola Normale Superiore,
             Piazza dei Cavalieri 7, 56126 Pisa, Italy.}
\author{Matteo Ambrosetti}
\affiliation{Scuola Normale Superiore,
             Piazza dei Cavalieri 7, 56126 Pisa, Italy.}
\author{Chiara Cappelli}
\affiliation{Scuola Normale Superiore,
             Piazza dei Cavalieri 7, 56126 Pisa, Italy.}
\email{chiara.cappelli@sns.it}

\title[]
  {On the Calculation of IR Spectra with a Fully Polarizable QM/MM Approach Based on Fluctuating Charges and Fluctuating Dipoles}

\begin{document}

\begin{abstract}
The fully polarizable QM/MM approach based on fluctuating charges and fluctuating dipoles, named QM/FQF$\mu$ (J. Chem. Theory Comput.  \textbf{2019}, \textit{15}, 2233-2245), is extended to the evaluation of nuclear gradients and the calculation of IR spectra of molecular systems in condensed phase. To this end, analytical equations defining first and second energy derivatives with respect to nuclear coordinates are derived and discussed. The potentialities of the approach are shown by applying the model to the calculation of IR spectra of Methlyoxirane, Glycidol and Gallic Acid in aqueous solution. The results are compared with the continuum QM/PCM and the polarizable QM/FQ, which is based on Fluctuating Charges only. %The results show a good agreement with experiments for all investigated moieties, thus showing the accuracy of the novel approach.
\end{abstract}

\newpage

\section{Introduction}

Vibrational spectroscopy, in particular infrared spectroscopy, is one of the most common techniques to study structural and dynamical features of molecular systems. Experimental spectra can be affected by a combination of effects, ranging from anharmonicity to solvent effects, the latter playing a relevant role because most experiments are conducted in the condensed phase.\cite{cappelli2011towards,bloino2012second,bloino2015anharmonic,mennucci2005model,mennucci1999analytical,cossi1998analytical} %For such a reason, computational chemistry has become a powerful tool to decipher and interpret IR spectra.\cite{barone2011computational} 
In this work, we especially focus on the development of a method to account for the mutual interaction between a molecular systems and its environment and its effect on the prediction of IR spectra. In fact, the presence of the environment can alter the electronic response of the target molecule to the external electric field and the vibrational frequecies associated with the normal modes. Therefore, approaches able to accurately describe environmental effects are required to obtain computed spectra directly comparable with experiments. 

In the computational practice, the effects of the environment on a given spectral property are usually included by resorting to focused models,\cite{warshel1976theoretical,warshel1972calculation,
miertuvs1981electrostatic,tomasi1994molecular,orozco2000theoretical,tomasi2005} which are based on the assumption that the spectral signal is essentially due to the target molecule (i.e. the solute in case of solutions) and the environment (i.e. the solvent) only modifies but not determines it. 

%By In case of strongly solute-solvent interactions, as in the case of Hydrogen Bonds (HB), the atomistic description of the solvent, rather than adopting a blurred continuum description of the environment, is crucial to construct a reliable theoretical model  \cite{miertuvs1981electrostatic,orozco2000theoretical,tomasi2005,tomasi1994molecular,mennucci2007continuum,mennucci2010continuum,
%mennucci2013modeling,lipparini2016perspective}. 

Besides the commonly used continuum solvation approaches \cite{mennucci2013modeling,lipparini2016perspective} the family of QM/MM methods,\cite{senn2009qm,lin2007qm,warshel1976theoretical,warshel1972calculation} may be exploited. %in which the solute is treated at the Quantum Mechanical (QM) level, whereas the environment is described by means of a molecular mechanics (MM) force field. 
Their quality is connected to the specific force field which is exploited to treat the MM portion, and on the approach which is used to define the QM/MM interaction. The latter can be modeled by means of the basic mechanical embedding approach or by resorting to the so-called electrostatic embedding, where the coupling between QM and MM portions is described in terms of the Coulomb law, i.e. the electrostatic interaction between the potential of the QM density and the fixed charges which are placed at MM atoms. This picture is refined in the so-called polarizable QM/MM approaches, in which the mutual QM/MM polarization is taken into account; it can be modeled in different ways, e.g. by resorting to distributed multipoles,\cite{day1996effective,kairys2000qm,mao2016assessing,loco2016qm,loco2018modelingijqc} induced dipoles,\cite{thole1981molecular,steindal2011excitation,jurinovich2014fenna,loco2018modeling} Drude oscillators\cite{boulanger2012solvent}, Fluctuating Charges (FQ)\cite{rick1994dynamical,rick1996dynamical,cappelli2016integrated} and the recently developed approach based on both Fluctuating Charges and Fluctuating Dipoles (FQF$\mu$).\cite{giovannini2019fqfmu}

In QM/FQF$\mu$ the MM portion is described in terms of both fluctuating charges and fluctuating dipoles, which are place at MM atoms positions and can vary as a response to the QM electric potential (the FQs) and MM atomic electronegativities and QM electric field (the F$\mu$s), respectively. Such an approach is a pragmatical extension of the QM/FQ approach, previously developed by some of the present authors, \cite{giovannini2018hyper,cappelli2016integrated,lipparini2012linear,giovannini2018hyper,lipparini2013gauge,
giovannini2017polarizable,giovannini2016effective,giovannini2018effective} in which the MM portion is described by means of electric charges which can be polarized by the QM density and viceversa. As a consequence, QM/FQF$\mu$ takes into account both the out-of-plane and anisotropy contributions to polarization thanks to the inclusion of the electric dipoles in the MM portion. It is worth remarking that similar approaches have been proposed in other contexts,\cite{stern1999fluctuating,naserifar2017polarizable,oppenheim2018extension,
mayer2007formulation,jensen2009atomistic,rinkevicius2014hybrid} however they are not based on a variational formalism and therefore are not specifically intended to model molecular properties/spectra. Also, to the best of our knowledge, they have never been extended to the calculation of molecular properties/spectra determined by the nuclear response to external fields.

The quality of QM/FQF$\mu$ at predicting electrostatic interaction energies has been recently discussed \cite{giovannini2019fqfmu} and some of the present authors have recently extended this approach to the calculation of electronic vertical transition energies of organic molecules in solution at the TD-DFT level.\cite{giovannini2019fqfmulinear} In this work, QM/FQF$\mu$ is further extended to the calculation of IR spectra, through its extension to energy nuclear derivatives. Remarkably, other QM/MM approaches have been extended to the calculation of energy gradients,\cite{caprasecca2014geometry,Cui_JCP_QMMMProp,field1990combined,Li_JCP_EFPPCMgrad,thompson1996qm,loco2017hybrid,kratz2016lichem,
dziedzic2016tinktep} however the only previous polarizable QM/MM approach extended to vibrational spectroscopy is the QM/FQ method developed by some of us.\cite{giovannini2018effective,giovannini2016effective,giovannini2017polarizable,lipparini2012analytical} 
 
The manuscript is organized as follows. In the next section, the FQF$\mu$ force field is presented and its coupling with a QM wavefunction defined at the SCF level (QM/FQF$\mu$) is detailed. Equations for analytical first and second energy derivatives are then presented and discussed. After a brief section discussing on the computational protocol which is adopted, numerical results are presented. In particular, QM/FQF$\mu$ is challanged against the description of IR spectra of three organic molecules in aqueous solution, namely methyloxirane, glycidol and gallic acid. IR spectra are computed by exploiting a hyerarchy of polarizable embedding approaches,  namely QM/PCM, QM/FQ and QM/FQF$\mu$. Computed spectra are compared to their experimental counterparts, which are taken from the literature.\cite{cappelli2005environmental,yang2009probing,kawiecki.1991} From the comparison between the different polarizable approaches, differences arising from the moving from continuum PCM to QM/FQ and QM/FQF$\mu$ fully atomistic approaches are highlighted, and the role of fluctuating charges, and their coupling with fluctuating dipoles is also appreciated. Some drawn conclusions and the discussion on future perspectives of this approach end the manuscript.

\section{Theoretical Model}

\subsection{QM/FQF$\mu$ Approach}

In the FQF$\mu$ force field each MM atom is endowed with both a charge $q$ and an atomic dipole $\bm{\mu}$, that can vary according to the external electric potential and electric field. Both charges and dipoles are described as s-type gaussian distribution functions:

\begin{align}
\rho_{q_i}(\gr{r}) & = \frac{q_i}{\pi^{\frac{3}{2}} R_{q_i}^3} \exp \left( -\frac{\abs{\gr{r}-\gr{r}_i}^2}{R_q^2}\right) \nonumber \\
\rho_{\bm{\mu}_i}(\gr{r}) & =\frac{\abs{\bm{\mu}_{i}}}{\pi^{\frac{3}{2}}R_{\bm{\mu}_i}^{3}} \hat{\gr{n}}_i\cdot\boldsymbol{\nabla} \left[\exp \left(-\frac{\left|\gr{r}-\gr{r}_{i}\right|^{2}}{R_{\bm{\mu}_i}^{2}} \right)\right]
\label{eq:distributioni}
\end{align}

where $R_{q_i}$ and $R_{\bm{\mu}_i}$ are the width of the Gaussian distributions $\rho_{q_i}$ and $\rho_{\bm{\mu}_i}$, respectively.  $\hat{\gr{n}}_{i}$ is a unit vector pointing to the dipole direction $\bm{\mu}_{i}$.

The total energy $E$ associated with a distribution of charges and dipoles can be written as: \cite{mayer2007formulation,giovannini2019fqfmu}

\begin{align}
\mathcal{E}(\mathbf{q},\bm{\mu}) & = \sum_{i}q_{i}\chi_{i} +  \frac{1}{2}
\sum_{i}q_{i}\eta_{i}q_{i} + \frac{1}{2}\sum_{i}\sum_{j\neq i} q_{i}\text{T}_{ij}^{qq} q_{j} + \sum_{i}\sum_{j\neq i}q_{i}\gr{T}_{ij}^{q\mu}\bm{\mu}_{j} + \nonumber \\
& + \frac{1}{2}\sum_{i}\sum_{j\neq i}\bm{\mu}_{i}^{\dagger}\mathbf{T}_{ij}^{\mu\mu}\bm{\mu}_{j} -\frac{1}{2}\sum_{i}\bm{\mu}_{i}^{\dagger}\alpha_{i}^{-1}\bm{\mu}_{i} 
\label{eq:energy-iniziale}
\end{align}

where $\chi$ is the atomic electronegativity, $\eta$ the chemical hardness and $\alpha$ the atomic polarizability. $\text{T}_{ij}^{qq}$, $\mathbf{T}_{ij}^{q\mu}$ and $\mathbf{T}_{ij}^{\mu\mu}$ are the charge-charge, charge-dipole and dipole-dipole interaction kernels, respectively.
If the gaussian distributions in Eq. \ref{eq:distributioni} are adopted, the functional form of the interaction kernels provided by Mayer \cite{mayer2007formulation} can be exploited (see also Ref. \cite{giovannini2019fqfmu}):

\begin{align}
\text{T}_{ij}^{qq} & =\frac{1}{|\gr{r}_{ij}|}\text{erf}\left(\frac{\abs{\gr{r}_{ij}}}{R_{q_i-q_j}}\right) \label{eq:kernelqq} \\
\text{\gr{T}}_{ij}^{q\mu} & =-\frac{\gr{r}_{ij}}{|\gr{r}_{ij}|^{3}}\left[\text{erf}\left(\frac{|\gr{r}_{ij}|}{R_{q_i-\mu_j}}\right)-\frac{2|\gr{r}_{ij}|}{\sqrt{\pi}R_{q_i-\mu_j}}\exp{\left({\frac{-|\gr{r}_{ij}|}{R_{q_i-\mu_j}}}\right)^{2}}\right] \label{eq:Tqmu} \\
{\mathbf{T}_{ij}^{\mu\mu}} & -= \frac{3\gr{r}_{i,j}\otimes\gr{r}_{i,j}-|\gr{r}_{i,j}|^{2}\mathbf{I}}{|\gr{r}_{i,j}|^{5}}\left[\text{erf}\left(\frac{|\gr{r}_{ij}|}{R_{\mu_i-\mu_j}}\right)-\frac{2}{\sqrt{\pi}}\frac{|\gr{r}_{ij}|}{R_{\mu_i-\mu_j}}\exp \left(-\frac{\gr{r}_{ij}}{R_{\mu_i-\mu_j}}\right)^{2}\right] + \nonumber \\
& + \frac{4}{\sqrt{\pi}{R_{\mu_i-\mu_j}^{3}}}\frac{\gr{r}_{ij}\otimes\gr{r}_{ij}}{|\gr{r}_{ij}|^{2}}\exp \left(-\frac{|\gr{r}_{ij}|}{R_{\mu_i-\mu_j}}\right)^{2}
\label{eq:Tmumu}
\end{align} 

In order to collect all quadratic terms in the charges, the diagonal elements of $\gr{T}^{qq}$ can be imposed to be equal to the atomic chemical hardness $\eta$, so that the width of the charge distribution $R_q$ is defined without the need of any parametrization.\cite{giovannini2019fqfmu} The same holds for the diagonal elements of $\gr{T}^{\mu\mu}$ and the dipole distribution $R_\mu$, which can be defined in terms of the atomic polarizabilities ($\alpha$).\cite{giovannini2019fqfmu} The definition of the gaussian width $R_{q_i}$ and $R_{\mu_i}$ in terms of $\eta_i$ and $\alpha_i$ limits the number of parameters which enter the definition of FQF$\mu$ to electronegativity, chemical hardness and polarizability for each atom type. 
Therefore, Eq. \ref{eq:energy-iniziale} can be rewritten as:

\begin{align}
\mathcal{E}(\mathbf{q},\bm{\mu}) & = \frac{1}{2}\sum_{i}\sum_{j} q_{i}{\text{T}_{ij}^{qq}}q_{j} +\frac{1}{2}\sum_{i}\sum_{j}\bm{\mu}_{i}^{\dagger}\mathbf{T}_{ij}^{\mu\mu}\bm{\mu}_{j} + \sum_{i}\sum_{j}q_{i}\textbf{T}_{ij}^{q\mu}\bm{\mu}_{j}^{\dagger} + \sum_{i}q_{i}\chi_{i} = \nonumber\\
& = \frac{1}{2}\mathbf{q}^{\dagger}\mathbf{T}^{qq}\mathbf{q} + \frac{1}{2}\bm{\mu}^{\dagger}\mathbf{T}^{\mu\mu}\bm{\mu} + \mathbf{q}^{\dagger}\mathbf{T}^{q\mu}\bm{\mu}+\bm{\chi}^{\dagger}\mathbf{q}
\label{eq:energiaMM_mat}
\end{align}
where a matrix notation has been adopted. In Eq. \ref{eq:energiaMM_mat}, charges are not forced to any value by external constraints and the equilibrium condition is reached when the Electronegativity Equalization Principle (EEP) is satisfied. %Such a principle states that at equilibrium each atom has the same electronegativity. 
If each MM molecule is constrained to assume a fixed, total charge equal to
Q$_\alpha$, Eq. \ref{eq:energiaMM_mat} can be written by exploiting a set of Lagrangian multipliers ($\lambda_\alpha$), whose number is equal to the total number of molecules in the MM portion.

\begin{align}
\mathcal{E}\left(\mathbf{q},\bm{\mu},\lambda_{\alpha}\right) & = E\left(\mathbf{r},\mathbf{q},\bm{\mu}\right) + \sum_{\alpha}\left[\lambda_{\alpha} \sum_{i} \left(q_{\alpha i}\right)-Q_{\alpha}\right] = \nonumber \\
& =\frac{1}{2}\sum_{i\alpha}\sum_{j\beta} q_{i\alpha}{\text{T}_{i\alpha,j\beta}^{qq}}q_{j\beta} +\frac{1}{2}\sum_{i}\sum_{j}\bm{\mu}_{i\alpha}^{\dagger}\mathbf{T}_{i\alpha,j\beta}^{\mu\mu}\bm{\mu}_{j\beta} + \sum_{i}\sum_{j}q_{i\alpha}\textbf{T}_{i\alpha,j\beta}^{q\mu}\bm{\mu}_{j\beta}^{\dagger} + \nonumber \\
& + \sum_{i\alpha}q_{i\alpha}\chi_{i\alpha} + \sum_{\alpha}\lambda_{\alpha}\left[\sum_{i} q_{\alpha i}-Q_{\alpha}\right] = \nonumber \\
& = \frac{1}{2}\mathbf{q}^{\dagger}\mathbf{T}^{qq}\mathbf{q} + \frac{1}{2}\bm{\mu}^{\dagger}\mathbf{T}^{\mu\mu}\bm{\mu} + \mathbf{q}^{\dagger}\mathbf{T}^{q\mu}\bm{\mu}+\bm{\chi}^{\dagger}\mathbf{q} + \bm{\lambda}^\dagger \gr{q}  
\label{eq:energia_finale_MM}
\end{align}

where $\alpha$ and $\beta$ run over molecules and the constraints $\lambda_{\alpha}$ are meant to preserve the total charge $Q_\alpha$ of each molecule. Therefore, the constrained minimum is found by imposing all derivatives of $\mathcal{F}$ with respect all variables to be equal to zero, thus resulting in the following linear system\cite{giovannini2019fqfmu}:
\begin{equation}
\left(
\begin{array}{ccc}
\mathbf{T}^{qq} & \mathbf{1}_{\bm{\lambda}} & \mathbf{T}^{q\mu} \\ 
\mathbf{1}^{\dagger}_{\bm{\lambda}} & \mathbf{0} & \mathbf{0}   \\
-\mathbf{T}^{q\mu^{\dagger}} & \mathbf{0} & \mathbf{T}^{\mu\mu}
\end{array}
\right)
\left({\begin{array}{c} 
\mathbf{q}\\
\bm{\lambda}\\ 
\bm{\mu} 
\end{array}}\right)
=
\left(\begin{array}{c} -\bm{\chi} \\ 
\mathbf{Q} \\
\mathbf{0}
\end{array}\right)
\qquad \Rightarrow \qquad
\mathbf{D}\mathbf{L}_{\lambda} = -\mathbf{C}_Q
\label{eq:MMlinearsys}
\end{equation}

where $\gr{1}_{\bm{\lambda}}$ is a rectangular matrix which accounts for the Lagrangians. $\mathbf{C}_Q$ is a vector containing atomic electronegativities and total charge constraints, whereas $\mathbf{L}_\lambda$ is a vector containing charges, dipoles and Lagrange multipliers.

FQF$\mu$ can be effectively coupled to a QM SCF wavefunction in a QM/MM framework. The QM density interacts as a classical density of charge with both charges and dipoles:

\begin{equation}
E_{QM/MM}= \sum_i V[\rho_{QM}](\gr{r}_i){q_i} - \bm{\mu}_i^{\dagger}\mathbf{E}[\rho_{QM}](\gr{r}_i)
\end{equation}

where $V[\rho_{QM}](\gr{r}_i)$ and $\mathbf{E}[\rho_{QM}](\gr{r}_i)$ are the electric potential and electric field, respectively, calculated at the $i$-th charge and $i$-th dipoles placed at $\gr{r}_i$.
Finally, the global QM/MM energy functional for a SCF-like description of the QM portion finally reads:
 
\begin{align}
\mathcal{E}(\mathbf{P},\mathbf{q},\bm{\mu},\bm{\lambda}) & = \text{tr} \bf{hP}+\frac{1}{2}\text{tr}\mathbf{PG(P)} + \frac{1}{2}\mathbf{q}^{\dagger}\mathbf{T}^{qq}\mathbf{q} + \frac{1}{2}\bm{\mu}^{\dagger}\mathbf{T}^{\mu\mu}\bm{\mu} + \mathbf{q}^{\dagger}\mathbf{T}^{q\mu}\bm{\mu}+\bm{\chi}^{\dagger}\mathbf{q} + \bm{\lambda}^\dagger \gr{q} + \nonumber \\
& + \mathbf{q}^{\dagger}\mathbf{V}(\mathbf{P}) - \bm{\mu}^{\dagger}\mathbf{E}(\mathbf{P})
\label{eq:funz_qmmm_fqfmu}
\end{align}

where $\mathbf{h}$ and $\mathbf{G}$ are the usual one- and two-electron matrices. The effective Fock matrix is defined as the derivative of the energy with respect to the density matrix:
\begin{equation}
 \label{eq:Fock_fqfmu}
 \tilde{F}_{\mu\nu} = \parz{\mathcal{E}}{P_{\mu\nu}} = h_{\mu\nu} + G_{\mu\nu}(\mathbf{P}) + \mathbf{V}^{\dagger}_{\mu\nu}\mathbf{q} - \mathbf{E}_{\mu\nu}^{\dagger}\bm{\mu}
\end{equation}
where the interaction of the electron density with both charges and dipoles are included through the coupling electrostatic terms.
Charges and dipoles are obtained by imposing the global functional to be stationary with respect to charges, dipoles and Lagrangian multipliers. 
\begin{equation}
 \label{eq:QMlinearsystem}
\left(
\begin{array}{ccc}
\mathbf{T}^{qq} & \mathbf{1}_{\bm{\lambda}} & \mathbf{T}^{q\mu} \\ 
\mathbf{1}^{\dagger}_{\bm{\lambda}} & \mathbf{0} & \mathbf{0}   \\
-\mathbf{T}^{q\mu^{\dagger}} & \mathbf{0} & \mathbf{T}^{\mu\mu}
\end{array}
\right)
\left({\begin{array}{c} 
\mathbf{q}\\
\bm{\lambda}\\ 
\bm{\mu} 
\end{array}}\right)
=
\left(\begin{array}{c} 
-\bm{\chi} \\ 
\textbf{Q}_{\text{tot}} \\
\mathbf{0}
\end{array}\right)
+
\left(\begin{array}{c}
-\gr{V}(\gr{P}) \\
\textbf{0} \\
\gr{E}(\gr{P})
\end{array}
\right)
\qquad \Rightarrow \qquad
\mathbf{D}\mathbf{L}_{\lambda} = -\mathbf{C}_Q - \mathbf{R}(\mathbf{P})
\end{equation}
Notice that, with respect to Eq. \ref{eq:MMlinearsys}, a new source term $\mathbf{R}(\mathbf{P}$), due to the coupling of both charges and dipole with the SCF density, arises. Again, $\mathbf{L}_{\lambda}$ is a vector containing charges, dipoles and Lagrangian multipliers. 

\subsection{Analytical Energy Derivatives}

In this section QM/FQF$\mu$ analytical first and second energy derivatives are presented and discussed. The following equations are defined in the so-called Partial Hessian Vibrational Approach (PHVA),~\cite{jin.1994,calvin.1996,biancardi.2012}  which has been amply exploited to treat vibrational phenomena of complex systems.\cite{lipparini2012analytical,giovannini2016effective,giovannini2017polarizable,giovannini2018effective} Within such a framework, it is assumed that the geometrical perturbation only acts on the QM portion of the system, whereas MM atoms are unaffected. Remarkably, the PHVA is fully consistent with a focused approach. For the sake of completeness, however, equations for first and second derivatives of FQF$\mu$ MM atoms are given in the Appendix section.
The following derivation directly follows what already reported by some of the present authors in case of QM/FQ.\cite{lipparini2012analytical} This allows to directly identify the additional terms which depend on the presence of fluctuating dipoles in the MM portion.

\subsubsection{Energy first derivatives}

The energy first derivative of Eq. \ref{eq:funz_qmmm_fqfmu} with respect to the $x$ nuclear displacement can be expressed by means of the chain rule:\cite{frisch1990,lipparini2012analytical}
\[
 \mathcal{E}^{x}(\mathbf{P},\mathbf{q},\bm{\mu},\bm{\lambda}) =  \frac{\partial \mathcal{E}}{\partial x} + \frac{\partial \mathcal{E}}{\partial \mathbf{P}}\frac{\partial \mathbf{P}}{\partial x} + \frac{\partial \mathcal{E}}{\partial \mathbf{q}}\frac{\partial \mathbf{q}}{\partial x} + 
 \frac{\partial \mathcal{E}}{\partial \bm{\mu}}\frac{\partial \bm{\mu}}{\partial x}+ \frac{\partial \mathcal{E}}{\partial \bm{\lambda}}\frac{\partial \bm{\lambda}}{\partial x}
\]
The last three terms vanish because of the stationarity conditions. The first term, which is the partial derivative of the energy with respect to the position of a QM nucleus, reads:
\begin{equation}
 \label{eq:pder1}
 \frac{\partial \mathcal{E}}{\partial x} = \mbox{tr}\ \mathbf{h}^x\mathbf{P} + \frac{1}{2}\mbox{tr}\ \mathbf{G}^{(x)}(\mathbf{P})\mathbf{P} + \mathbf{q}^\dagger\mathbf{V}^{(x)}(\mathbf{P}) - 	\bm{\mu}^{\dagger}\mathbf{E}^{(x)}(\mathbf{P})
\end{equation}
where
\begin{eqnarray}
V^{(x)}_i(\mathbf{P}) & = & \sum_{\mu\nu} P_{\mu\nu} V_{\mu\nu,i}^x + \mbox{nuclear\ contribution} \\
& = & \frac{Z_\zeta}{|\mathbf{R}_\zeta - \mathbf{r}_i|^2} - \sum_{\mu\nu} \Braket{\frac{\partial (\chi_\mu\chi_\nu)}{\partial \mathbf{R}_\zeta} | \frac{1}{\abs{\mathbf{r} - \mathbf{r}'}} | \delta(\mathbf{r}'-\mathbf{r}_i)} P_{\mu\nu} \\
E^{(x)}_i(\mathbf{P}) & = & \mathbf{\nabla}_{r_i} V^{(x)}_i(\mathbf{P})
\end{eqnarray}

The term involving the derivatives of the density matrix can be computed starting from the idempotency condition: 
\[
 -\mathbf{P}\tilde{\mathbf{F}}\mathbf{P}\mathbf{S}^x_{oo} = - \tilde{\mathbf{W}}\mathbf{S}^x_{oo}
\]
where the subscript $oo$ denotes the occupied--occupied block of the matrix in the MO basis, and $\mathbf{W}$ is the energy-weighted density matrix contribution. By collecting all the terms:

\begin{equation}
 \label{eq:1derQMMM}
  \mathcal{E}^{x}(\mathbf{P},\mathbf{q},\bm{\lambda}) = \mbox{tr}\ \mathbf{h}^x\mathbf{P} + \frac{1}{2}\mbox{tr}\ \mathbf{G}^{(x)}(\mathbf{P})\mathbf{P} + \mathbf{q}^\dagger \mathbf{V}^{(x)}(\mathbf{P}) - 	\bm{\mu}^{\dagger}\mathbf{E}^{(x)}(\mathbf{P}) - \mbox{tr}\mathbf{W}\mathbf{S}^x_{oo}
\end{equation}

Notice that the term ($\mathbf{q}^\dagger \mathbf{V}^{(x)}(\mathbf{P})$) is the same as computed in the QM/FQ approach.\cite{lipparini2012analytical} Therefore, the inclusion of fluctuating dipoles gives rise to the additional term $\bm{\mu}^{\dagger}\mathbf{E}^{(x)}(\mathbf{P})$.

\subsubsection{Energy second derivatives}

The energy second derivative with respect to nuclear displacements $x$ and $y$ is obtained by differentiating eq. \ref{eq:1derQMMM} and by exploiting once again the chain rule:
\begin{eqnarray}
 \label{eq:2der1}
  \mathcal{E}^{xy} & = & \sum_{\mu\nu} \left [ \hmn^{xy} + \frac{1}{2}\Gmn^{(xy)}(\bP) + \bq^\dagger\bVmn^{xy} - \bmu^\dagger\bEmn^{xy} \right ]\Pmn - \mbox{tr}\ \bW\bS^{xy}\nonumber - \mbox{tr}\ \bW^y\bS^x\\
           & + & \sum_{\mu\nu} \left [ \hmn^x + \Gmn^{(x)}(\bP) + \bq^\dagger \bVmn^x -\bmu^\dagger\bEmn^x\right ] \Pmn^y + \sum_{\mu\nu}\bL^{y\dagger} \bRmn^x\Pmn
\end{eqnarray}
Thus, the derivatives of the off-diagonal blocks of the density matrix and charges/dipoles need to be calculated. Density matrix derivatives can be obtained through a Coupled Perturbed Hartree--Fock (CPHF) or Kohn-Sham (CPKS) procedure. %Following Frisch and co-workers\cite{frisch1990}, and Lipparini et al. \cite{lipparini2012analytical}, we obtain:
%the CPHF equation can be obtained by differentiating once the SCF equations are in their Liouville form:

FQF$\mu$ charge and dipole derivatives can be calculated by differentiating eq. \ref{eq:QMlinearsystem}:
\begin{equation}
 \label{eq:qx}
 \bD\bL^x = - \bR^{(x)}(\bP) - \bR(\bP^x)
\end{equation}

%Substitution yields:
The Fock matrix derivative is defined as:
\begin{equation}
 \label{eq:Fx}
 \Fmnt^x =  \Fmnt^{(x)} +  \Gmn(\bP^x) - \bRmn^\dagger \bD^{-1} \bR(\bP^x)
\end{equation}
where $\Fmnt^{(x)}$, which collects all explicit derivatives of the Fock matrix, reads:
\[
 \Fmnt^{(x)} = \hmn^x + \Gmn^{(x)}(\bP) + \bL^\dagger\bRmn^x + \bRmn^\dagger \bL^{(x)}
\]

By rearranging the terms, the CPHF/CPKS equations are obtained (MO basis):

\begin{eqnarray}
 \label{eq:CPHF_MO1}
 \epsilon_i P^x_{ia} - \epsilon_a P^x_{ia} = -\tilde{Q}_{ia} & + & \sum_{jb} \left [ \langle aj || ib\rangle - \mathbf{R}_{ia}^\dagger\bD^{-1}\mathbf{R}_{jb} \right ] P^x_{jb} \nonumber \\
 & + & \sum_{jb} \left [ \langle ab || ij\rangle - \mathbf{R}_{ia}^\dagger\bD^{-1}\mathbf{R}_{bj} \right ] P^x_{bj}
\end{eqnarray}

By taking the adjunct equation and introducing the following matrices (we assume orbitals to be real):
\begin{equation}
 \label{eq:Amat2}
 \tilde{A}_{ia,jb} =  (\epsilon_a - \epsilon_i)\delta_{ij}\delta_{ab} + \langle aj || ib\rangle - \mathbf{R}^\dagger_{ia}\bD^{-1}\mathbf{R}_{jb}
\end{equation}

\begin{equation}
 \label{eq:Bmat2}
 \tilde{B}_{ia,jb} =  \langle ab || ij\rangle  - \mathbf{R}_{ia}^\dagger\bD^{-1}\mathbf{R}_{bj}
\end{equation}
the following equation is obtained:
\begin{equation}
 \label{eq:CPHFFQCasida}
\left (
 \begin{array}{cc}
  \tilde{\mathbf{A}} & \tilde{\mathbf{B}} \\
  \tilde{\mathbf{B}}^* & \tilde{\mathbf{A}}^* \\
 \end{array}
\right )
\left (
 \begin{array}{c}
  \mathbf{X} \\
  \mathbf{Y} \\
 \end{array}
\right )
=
\left (
 \begin{array}{c}
  \mathbf{Q} \\
  \mathbf{Q}^* \\
 \end{array}
\right )
\end{equation}

where
\begin{equation}
 \label{eq:1-Amat2}
 \tilde{A}_{ia,jb} =  (\epsilon_a - \epsilon_i)\delta_{ij}\delta_{ab} + \langle aj || ib\rangle - \mathbf{R}^\dagger_{ia}\bD^{-1}\mathbf{R}_{jb}
\end{equation}
\begin{equation}
 \label{eq:1-Bmat2}
 \tilde{B}_{ia,jb} =  \langle ab || ij\rangle  - \mathbf{R}_{ia}^\dagger\bD^{-1}\mathbf{R}_{bj}
\end{equation}
\begin{equation}
 \label{eq:1-RHS}
 \tilde{Q}_{ia} = \tilde{F}_{ia}^{(x)} - G_{ia}(S^x_{oo}) - \tilde{\bF}S^{x}_{ia} + \mathbf{R}^\dagger_{ia}\bD^{-1}\bR(\bS^x_{oo})
\end{equation}

Therefore, the derivatives of the density matrix and FQF$\mu$ charge/dipole derivatives with respect to QM nuclear positions are obtained by solving Eqs. \ref{eq:CPHFFQCasida} and \ref{eq:qx}, respectively. Notice that such a derivation is coherent with what has been reported for linear response in the zero-frequency limit.\cite{giovannini2019fqfmulinear}

To summarize, FQF$\mu$ contributions to analytical second derivatives can be grouped into three categories:
\begin{enumerate}
 \item explicit terms:
\[
 \bq^\dagger \bV^{(xy)} - \bmu^\dagger \bE^{(xy)} + \bL^{(x)\dagger}\bR^{(y)} 
\]
 \item contributions to Fock matrix derivatives:
\[
 \bL^\dagger \bRmn^x + \bL^{(x)\dagger}\bRmn
\]
 \item additional terms to the CPHF/CPKS matrix:
\[
 - \mathbf{R}^\dagger_{ia}\bD^{-1}\mathbf{R}_{jb}
\]
\end{enumerate}

Notice that only the last term is required in case of electric perturbations.Also, similarly to what has already been discussed for energy first derivatives, QM/FQF$\mu$ second derivatives differ from QM/FQ ones because additional terms depending on fluctuating dipoles need to be included.\cite{lipparini2012analytical} 

\section{Computational Details}

R-methyloxirane (MOXY), (S)-Glycidol (GLY) and Gallic Acid (GA) geometries were optimized at the B3LYP/ aug-cc-pVDZ. Solvent effects on solutes' geometries were included through the Polarizable Continuum Model (PCM).  MOXY and GLY Molecular Dynamics (MD) simulations were carried out as detailed in previous works by some of the present authors\cite{lipparini2013optical,giovannini2018effective} by using GROMACS.\cite{Gromacs5} 

A 25 ns MD simulation of GA in aqueous solution was performed by using a similar computational protocol. A GA molecule was placed at the center of a cubic box and solvated with 6025 TIP3P water molecules.\cite{mark2001structure} The parameters used to describe the GA inter-/intra-molecular interactions were taken from the GAFF force field\cite{GAFF} by using the ANTECHAMBER package.\cite{antechamber} GA molecule was keeped fixed during all the steps of the simulation. To model intermolecular solute-solvent interactions,  ga partial charges were computed by relying on the Hirshfeld population analysis,\cite{CM5} as previously done by some of the present authors.\cite{giovannini2019simulating,giovannini2018hyper} Partial charges were computed at the B3LYP/6-311++G** level of theory by including solvent effects by means of PCM.
Two subsequent 100 ps runs were performed for equilibration purposes by using NVT and NPT ensembles, respectively. A 25 ns NVT MD simulation was then performed to sample the configurartional space at time steps of 1 fs and by saving coordinates every 10 ps. The system was simulated by using three-dimensional periodic boundary conditions; non-bonded interactions cutoff was set to 10 \AA. A particle mesh Ewald (PME) correction for the long-range electrostatics was applied and the temperature was maintained at 300 K  by using the velocity rescale algorithm.

A total of 200 uncorrelated snapshots were extracted from the last 20 ns, 50 ns and 25 ns of MD simulations in case of MOXY, GLY and GA, respectively. For  each snapshot, a  sphere  centered  at  the  solute's geometric center was cut.A cutting radius of 12 \AA\ was used for MOXY and GLY, whereas a cutting radius of 15 \AA was used for GA. 

QM/FQ and QM/FQF$\mu$ partial geometry optimization for the solute on each snapsho was performed according to the default settings of Gaussian16,\cite{gaussian16} by keeping all water molecules fixed. Finally, infrared (IR) spectra were calculated on each partially optimized snapshot with the QM/FQ and QM/FQF$\mu$ models at the B3LYP/aug-cc-pVDZ level (MOXY and GLY) and B3LYP/6-311++G** level (GA). SPC parameters for water \cite{rick1994dynamical} were used for FQ calculations. The set of parameters recently developed by som of us for QM/FQF$\mu$ calculations in aqueous solutions were exploited.\cite{giovannini2019fqfmu}. IR data were averaged to obtain final spectra for the three solutes; peaks were convoluted with a Lorentzian lineshape, with Full Width at Half Maximum (FWHM) of 4 $\mathrm{cm}^{-1}$. For the sake of comparison, QM/PCM IR spectra were also computed.

All QM/FQ and QM/FQF$\mu$ calculations were performed by using a locally modified version of the Gaussian 16 package,\cite{gaussian16} where QM/FQF$\mu$ analytical energy first derivatives were implemented.

\section{Numerical Results}

\begin{figure}[htbp!]
\centering
\subfloat[][\gr{a) MOXY}]{\includegraphics[width=.14\textwidth]{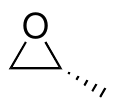}}\qquad
\subfloat[][\gr{b) GLY}]{\includegraphics[width=.25\textwidth]{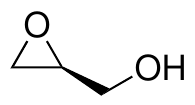}}\qquad
\subfloat[][\gr{c) GA}]{\includegraphics[width=.25\textwidth]{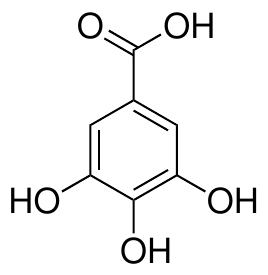}}
\caption{MOXY (\gr{a}), GLY (\gr{b}) and GA (\gr{c}) molecular structures.}
\label{fig:mol-struc}
\end{figure}

In this section, the results obtained by applying QM/FQF$\mu$ to the calculaiton of IR spectra of MOXY, GLY and GA in aqueous solutions are reported (see Fig. \ref{fig:mol-struc}, panels a-c for molecular structures). MOXY is a widely exploited test system for computational models\cite{barone2014accurate,hodecker2016simulation,ruud2005importance,tam2004coupled,mukhopadhyay2006solvent,
carnell1991experimental,science.2016}. GLY, which bears an additional hydroxil group, has been previously studied both theoretically and experimentally.\cite{giovannini2018effective,yang2009probing} In particular, it has been shown that eight different GLY conformers exist in aqueous solution, thus its theoretical modeling is challenging.\cite{sun2007dft,yang2009probing,giovannini2018effective}
GA is an organic acid characterized by the presence of three hydroxil groups linked to the aromatic ring. The modeling of IR spectra of GA in aqueous solution is also challenging, because it has been previously reported by one of the present authors that the implicit PCM fails at reproducing the experimental IR spectrum, and that experimental spectral features can be recovered by adopting a supramolecular approach which includes eight explicit water molecules in the QM portion.\cite{cappelli2005environmental} Therefore, it appears to be an ideal test-bed for QM/FQF$\mu$.

QM/FQF$\mu$ spectra will be compared to QM/FQ, which only considers fluctuating charges on MM atoms. The rerason for such a comparison is twofold: (i) QM/FQF$\mu$ is formally an extension of QM/FQ, and (ii) QM/FQ has been successfully applied to vibrational spectra of molecular systems in aqueous solution.\cite{giovannini2016effective,giovannini2017polarizable,giovannini2018effective}  Thus, the comparison between the two approaches can directly quantify the relevance of fluctuating dipoles in the description of vibrational spectra and report on the performance of the novel QM/FQF$\mu$ method.

\subsection{Methyloxirane in Aqueous Solution}

\begin{figure}[htbp!]
\centering
\includegraphics[width=.5\textwidth]{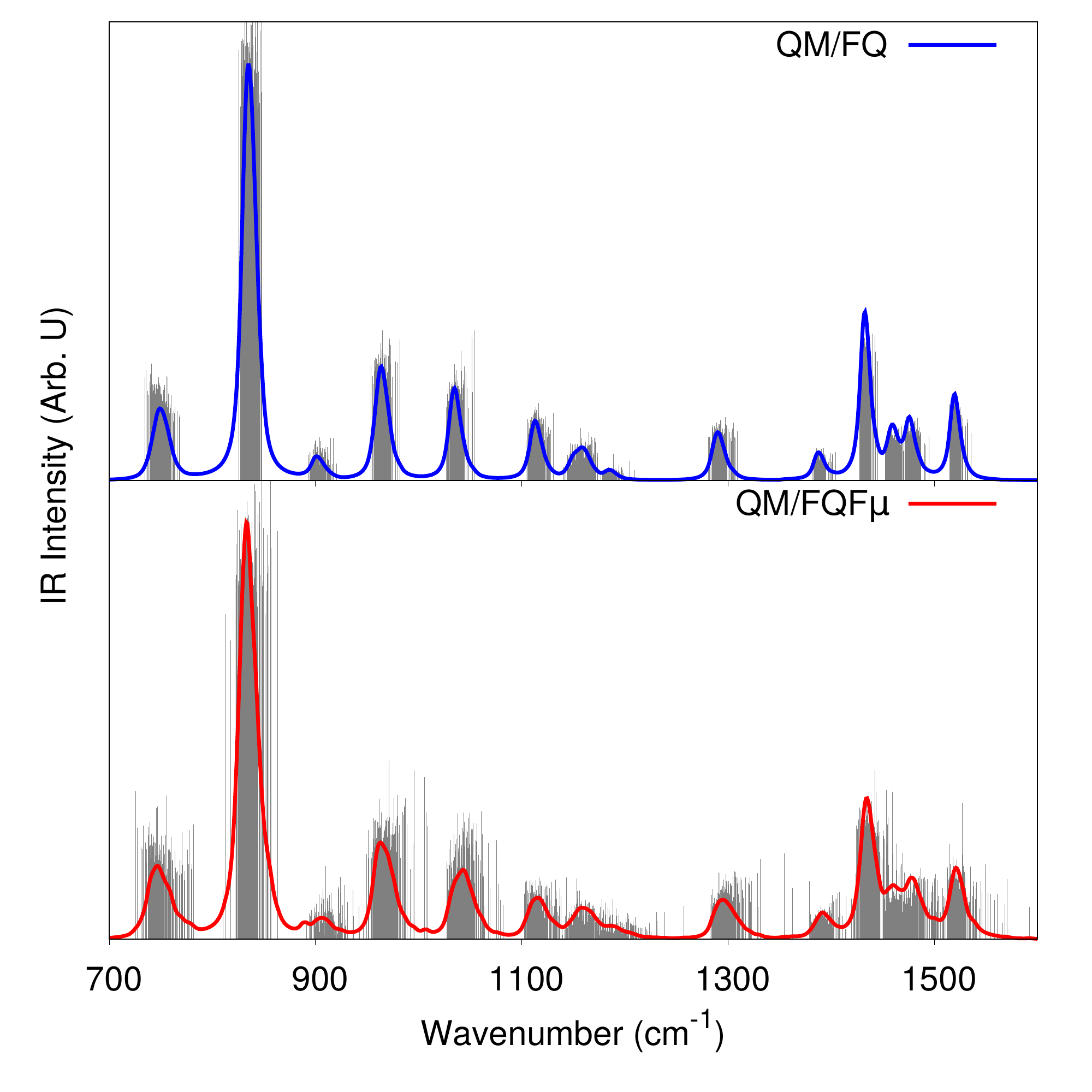}
\caption{Calculated QM/FQ (top) and QM/FQF$\mu$ (bottom) stick and convoluted IR spectra of MOXY in aqueous solution. FWHM = 4 cm$^{-1}$. %tommaso: che vuol direQUI!!!!!! CONTROLLA!!!!! i picchi a 900 con fqfmu
}
\label{fig:moxy-stick}
\end{figure}

%As a first test case, QM/FQF$\mu$ is applied to the IR spectrum of MOXY in aqueous solution. 
Figure \ref{fig:moxy-stick} compares QM/FQ (top) and QM/FQF$\mu$ (bottom) stick and convoluted spectra for 200 snapshots extracted from the MD simulation. Such a number of snapshots is enough to reach convergence.\cite{giovannini2017polarizable,giovannini2018effective} %differently to what it has been reported for Optical Rotation of the same system.\cite{lipparini2013optical}
Stick spectra are obtained by plotting raw data extracted from each frequency calculation. 
Figure \ref{fig:moxy-stick} clearly shows that both QM/FQ and QM/FQF$\mu$ IR wavenumbers and dipole strengths depend on the snapshot, i.e. on the arrangement of water molecules around the solute. As compared to QM/FQ, QM/FQF$\mu$ exhibits a larger spread, in particular in the vibrational wavenumbers. The largest variability of QM/FQF$\mu$ sticks occurs in the region between 1450 and 1500 cm$^{-1}$, which is associated to methyl and CH bending modes (see Figures S1 in the Supporting Information (SI) for a pictorial view of the normal modes for a randomly chosen snapshot). 

Clearly, band broadening is automatically considered in both QM/FQ and QM/FQF$\mu$ approaches coupled with the dynamical description given by the MD simulation, which samples over the solute-solvent phase-space. Therefore, solvent inhomogeneous broadening (due to the fluctuations of the solvent molecules) does not need to be artificially considered by imposing a pre-defined (and arbitrary) band-width, which is instead necessary when other static approaches, such as PCM, are used. The lorentzian convolution obtained by using a FWHM of 4 cm$^{-1}$ is plotted for both QM/FQ and QM/FQF$\mu$ spectra. Notice that QM/FQ IR spectrum is identical to what we reported in case of the three layer QM/FQ/PCM approach, where the PCM is used as third layer and coupled to both QM and FQ portions.%to impose non-periodic boundary conditions. 
Such a similarity means that water molecules explicitly included in snapshots are able to also account for bulk water effects.

\begin{figure}[htbp!]
\centering
\includegraphics[width=.5\textwidth]{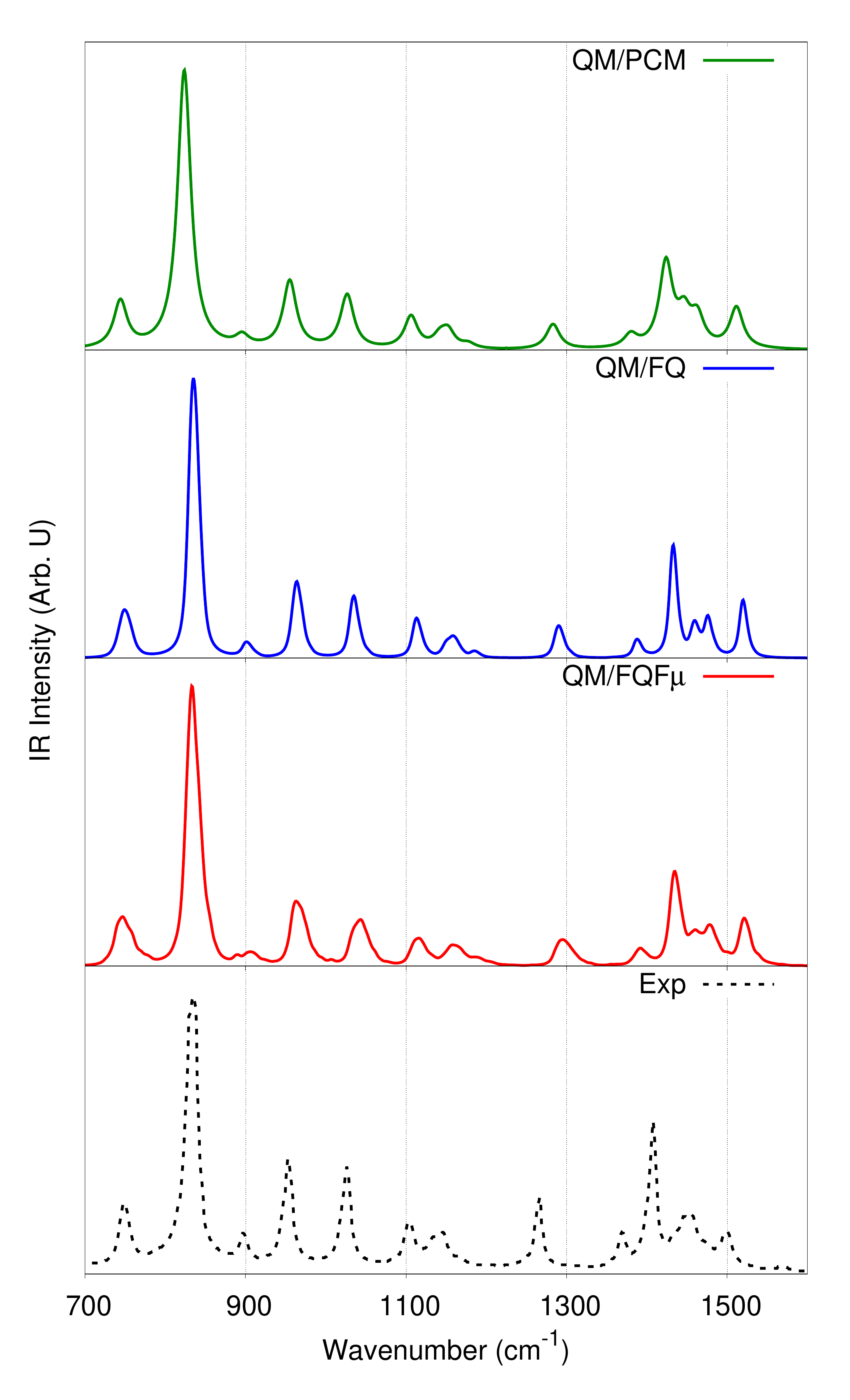}
\caption{Calculated QM/PCM (green), QM/FQ (blue) and QM/FQF$\mu$ (red) IR spectra of MOXY in aqueous solution (QM/PCM FWHM = 10 cm$^{-1}$; QM/MM FWHM = 4cm$^{-1}$). The experimental spectrum (dashed black) is reproduced from Ref. \citenum{kawiecki.1991}.}
\label{fig:moxy-ir}
\end{figure}

We now move to the comparison between computed and experimental spectra (see Fig.\ref{fig:moxy-ir}). Notice that the experimental spectrum was measured for neat liquid MOXY rather than aqueous solution;\cite{kawiecki.1991} therefore, some discrepancies with our computed results need to be expected. 

%In figure \ref{fig:moxy-ir}, QM/PCM, QM/FQ and QM/FQF$\mu$ IR spectra are compared to the experiment. 
The computed and experimental IR spectra are dominated by an intense band at about 850 cm$^{-1}$. This signal is given by the symmetric stretching of the C-O bond of the epoxyl group. QM/FQF$\mu$ band is blueshifted with respect to QM/FQ, thus the inclusion of fluctuating dipoles increases solute-solvent interactions. In fact, assuming that the normal modes are equally predicted by both QM/FQ and QM/FQF$\mu$ approaches, an increase in the solute-solvent interactions is reflected in a decrease in the solute intramolecular bond strengths, therefore, resulting in a blueshifted. Similar considerations apply also to the regions between 900-1100 cm$^{-1}$, where the normal modes involve vibrations of the MOXY oxygen atom. In the other regions of the spectra, such a blueshift is instead not recorded. This can be explained by the fact that the normal modes do not involve vibrations of the oxygen atom, which is the only MOXY atom potentially exhibiting an Hydrogen Bond with the solvent molecules.

Further differences between QM/FQ and QM/FQF$\mu$ are computed for the inhomogeneous band broadening, which is almost absent in case of QM/FQ, whereas it affects almost all QM/FQF$\mu$ bands. This is not unexpected if the raw data depicted in Fig. \ref{fig:moxy-stick} are considered. In fact, as already pointed out, QM/FQF$\mu$ generally spreads a larger energy region for each vibrational normal mode. 

%QUI!!!!!!! QUESTA FRASE FINO A GLY VA RIVISTA PERCHE' NON CHIARA However, by also referring to QM/PCM results, all the considered approaches to include environmental effects give similar IR spectra, thus showing that the presence of water is not significantly alters MOXY vibrational properties. This is confirmed by the fact that our computed results are nicely in agreement with the experiment, although the latter is measured for liquid neat MOXY. 
The major discrepancies with the experiment are reported in case of QM/FQ and QM/FQF$\mu$ in case of the inhomogenous band broadening, which is almost absent in the experiment. Thus, the main feature added by aqueous solution seem to be a larger broadening of vibrational bands. %Notice that this was not reported in case of Raman spectra.\cite{giovannini2017polarizable}

\subsection{Glycidol in Aqueous Solution}

\begin{figure}[htbp!]
\centering
\includegraphics[width=.5\textwidth]{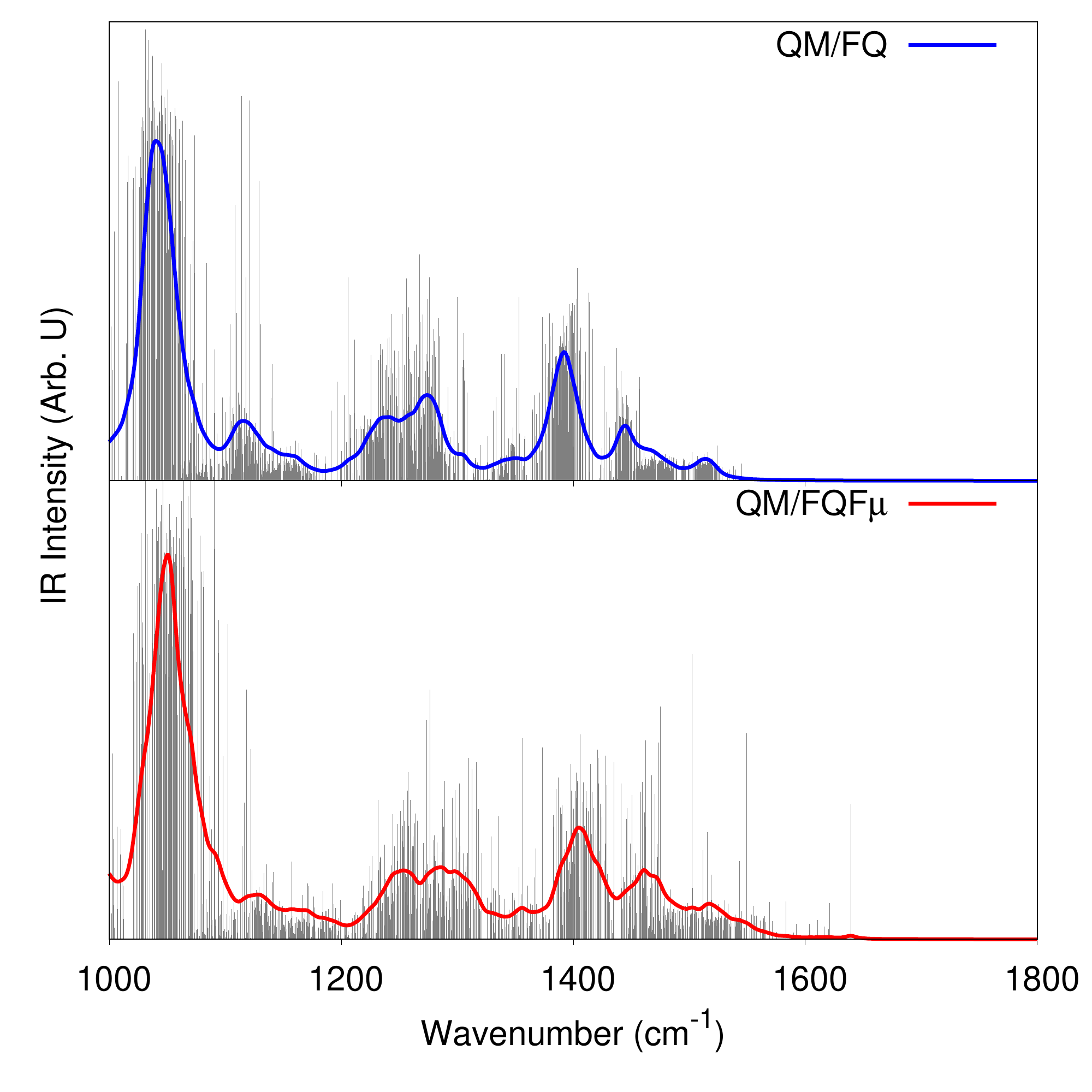}
\caption{Calculated QM/FQ (top) and QM/FQF$\mu$ (bottom) stick and convoluted IR spectra of GLY in aqueous solution (FWHM = 4 cm$^{-1}$).}
\label{fig:gly-stick}
\end{figure}

Similarly to MOXY, QM/FQ and QM/FQF$\mu$ IR spectra of GLY in aqueous solution were calculated by averaging over 200 snapshots extracted from the MD simulation.\cite{giovannini2018effective} QM/FQ and QM/FQF$\mu$ raw data are graphically plotted in Fig. \ref{fig:gly-stick}, together with their lorentzian convolution. 
As stated before, the case of GLY in aqueous solution is far more complicated than MOXY, because GLY is a flexible molecule, i.e. it exists in different conformations. This is reflected in the stick spectra depicted in Fig. \ref{fig:gly-stick}, which show a larger variability both in intensities and wavenumbers as compared to MOXY (see Fig. \ref{fig:moxy-stick}); this applies to both QM/FQ and QM/FQF$\mu$ calculations with the exception of the region around 1110 cm$^{-1}$. There, QM/FQ shows a substantial variation in intensity, whereas QM/FQF$\mu$ spreads a larger wavenumber range. The spreading of the intensities is instead larger for QM/FQF$\mu$ in the region between 1400-1600 cm$^{-1}$. Despite such discrepancies between QM/FQ and QM/FQF$\mu$, inhomogeneous broadening is described by both approaches.% coupled with the dynamical MD description of the solvation phenomenon. 

\begin{figure}[htbp!]
\centering
\includegraphics[width=.5\textwidth]{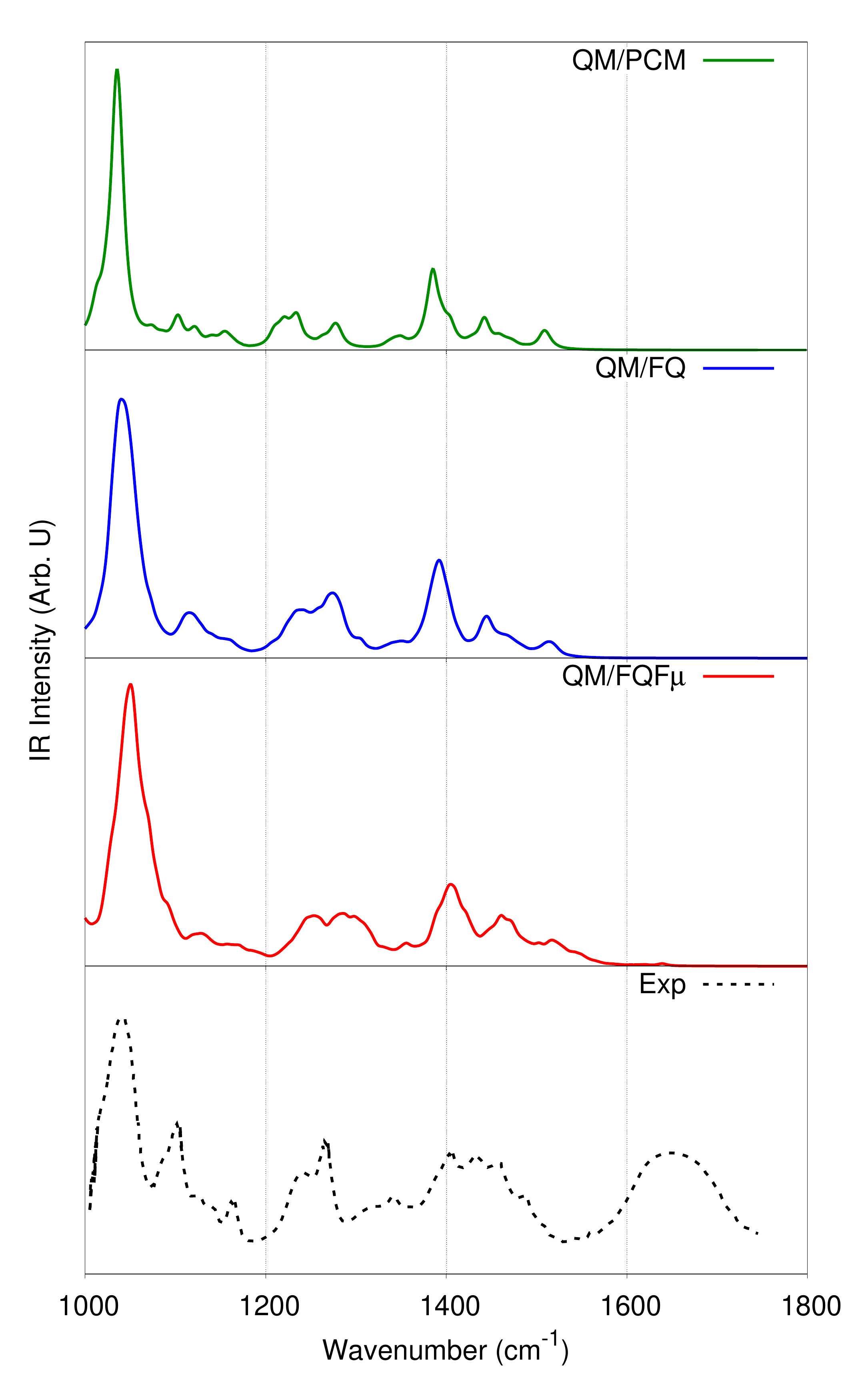}
\caption{Calculated QM/PCM (green), QM/FQ (blue) and QM/FQF$\mu$ (red) IR spectra of GLY in aqueous solution (QM/PCM FWHM = 10 cm$^{-1}$; QM/MM FWHM = 4cm$^{-1}$). The experimental spectrum (dashed black line) is reproduced from Ref. \citenum{yang2009probing}.}
\label{fig:gly-ir}
\end{figure}

QM/PCM, QM/FQ and QM/FQF$\mu$ convoluted IR spectra are compared to experiments in Fig. \ref{fig:gly-ir}. The experimental IR spectrum i reproduced from Ref. \citenum{yang2009probing}. Normal modes for a randomly chosen snapshot extracted from the MD are depicted in Figure S2 in the S for the region 700-1800 cm$^{-1}$ .

Computed and experimental IR spectra are dominated by an intense band at about 1050 cm$^{-1}$, assigned to a diffuse stretching/bending normal mode, involving the hydroxyl group. %All investigated approaches predict very similar frequencies for this band (???? QUI!!!!!!!! CHE VUOI DIRE???) However, 
QM/FQF$\mu$ spectrum is generally blueshifted with respect to both QM/FQ and QM/PCM, probably due to the fact it predicts larger solute-solvent interactions.% are predicted bigger. In fact, similar considerations that was applied in case of MOXY in aqueous solution, i.e. the most large is the solute-solvent interaction the more blue-shifted is a band, are still valid. The same analysis is valid also for all the other bands in the computed spectra, which are generally blushifted by going from a blurred continuum approach, such as QM/PCM, to the atomistic QM/FQ to QM/FQF$\mu$. 

Moving to the comparison with experimental data, both the experimental and computed spectra are characterized by a two-peak-shaped band between 1200 and 1300 cm$^{-1}$, which is assigned to the C-OH and C-CH bending modes (at about 1230 and 1270 cm$^{-1}$, respectively). Furthermore, above 1400 cm$^{-1}$ a three-peak-shaped band can be identified, due to the C-OH bending (1395 cm$^{-1}$), a diffuse C-CH bending (1440 cm$^{-1}$) and a CH$_{2}$ bending (1465 cm$^{-1}$). 

QM/FQ, QM/FQF$\mu$ and the experimental IR spectru are nicely in agreement. In fact, most of relative intensities and the band broadening are correctly reproduced. In particular, QM/FQ accurately predicts the two-peak band between 1200 and 1300 cm$^{-1}$, whereas QM/FQF$\mu$ is able to catch the inhomogeneity of the three-peak-shaped band between 1400 and 1500 cm$^{-1}$. Some discrepancies are reported in case of the peak at about 1100 cm$^{-1}$ (due to the CH scissoring), which is predicted to have a very low in intensity by both atomistic QM/MM approaches, whereas it is the second most intense peak in the experimental spectrum. Notice that it has been reported that the relative intensity of this peak can be correctly reproduced if a supermolecule approach is adopted,, i.e. if  water molecules are included in the QM region.\cite{yang2009probing} These findings, together with the results obtained by adopting our QM/classical modeling, suggest that the inclusion of non-electrostatic interactions, which are considered in the full QM supermolecule approach, can play a relevant role to improve the quality of the computed spectrum in this region. %tommaso: perche'? QUI!!!!!!!!!!!!!! NON CAPISCO BENE... SALTO LOGICO?
%Moreover, a minor deviation in the vibrational energies due to the DFT level of theory and to the lack of anharmonicity is reported. 
Overall, it is worth noticing that the continuum QM/PCM approach cannot reproduce the experimental spectrum (see top of Fig. \ref{fig:gly-ir}), thus remarking once again the huge potentialities of our atomistic QM/FQ and QM/FQF$\mu$ approaches to model vibrational spectra of solutes strongly interacting with the aqueous environment.

To end the discussion on the IR spectrum of GLY in aqueous solution, we point out that the broad band measured between 1600-1700 cm$^{-1}$ in the experiment is not reproduced by any of the selected QM/classical approaches. As already reported by some of the present authors \cite{giovannini2016effective,giovannini2018effective} and in Refs. \cite{losada2008solvation,losada2008lactic,yang2009probing}, such band is due to the OH bending mode of water molecules linked to GLY; therefore, it cannot be modeled by our approaches, in which the normal modes/frequencies of the environment are not computed.

\subsection{Gallic Acid in Aqueous Solution}

\subsubsection{MD Analysis}

\begin{figure}[htbp!]
\centering
\includegraphics[width=.2\textwidth]{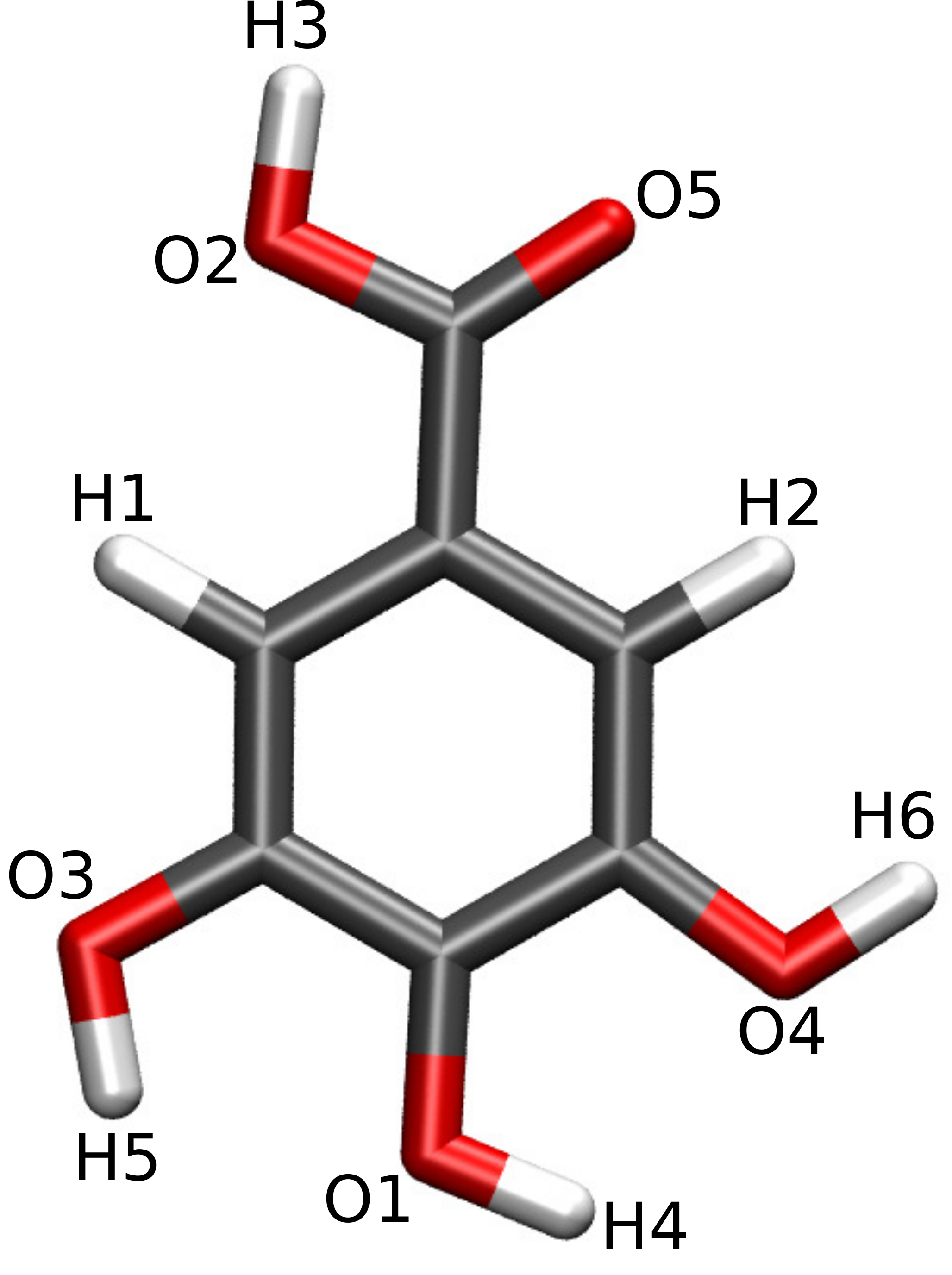}
\caption{Gallic Acid (GA) molecular structure with atom labelling.}
\label{fig:ga-struc}
\end{figure}
\begin{figure}[htbp!]
\centering
\includegraphics[width=.48\textwidth]{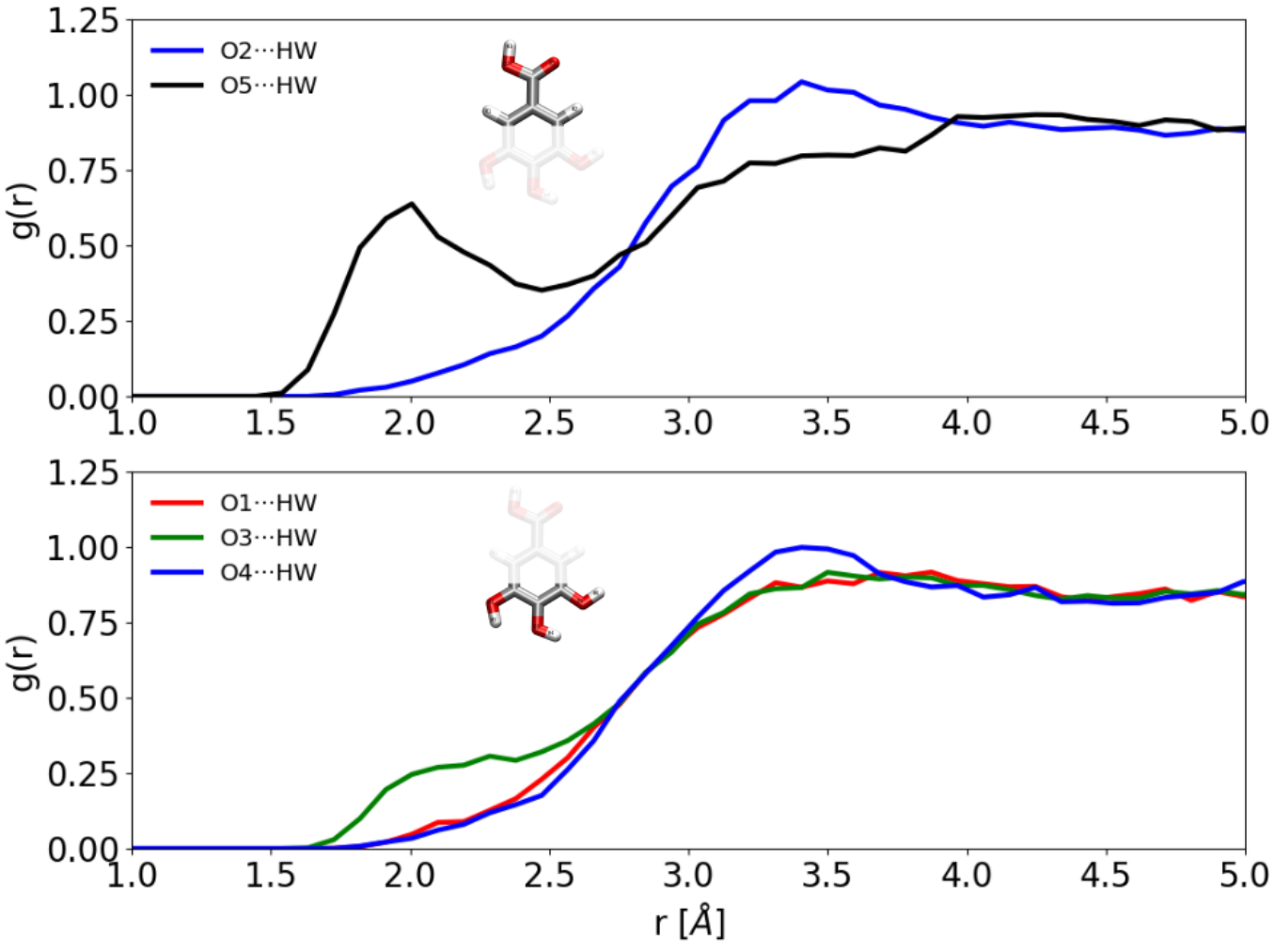}
\includegraphics[width=.48\textwidth]{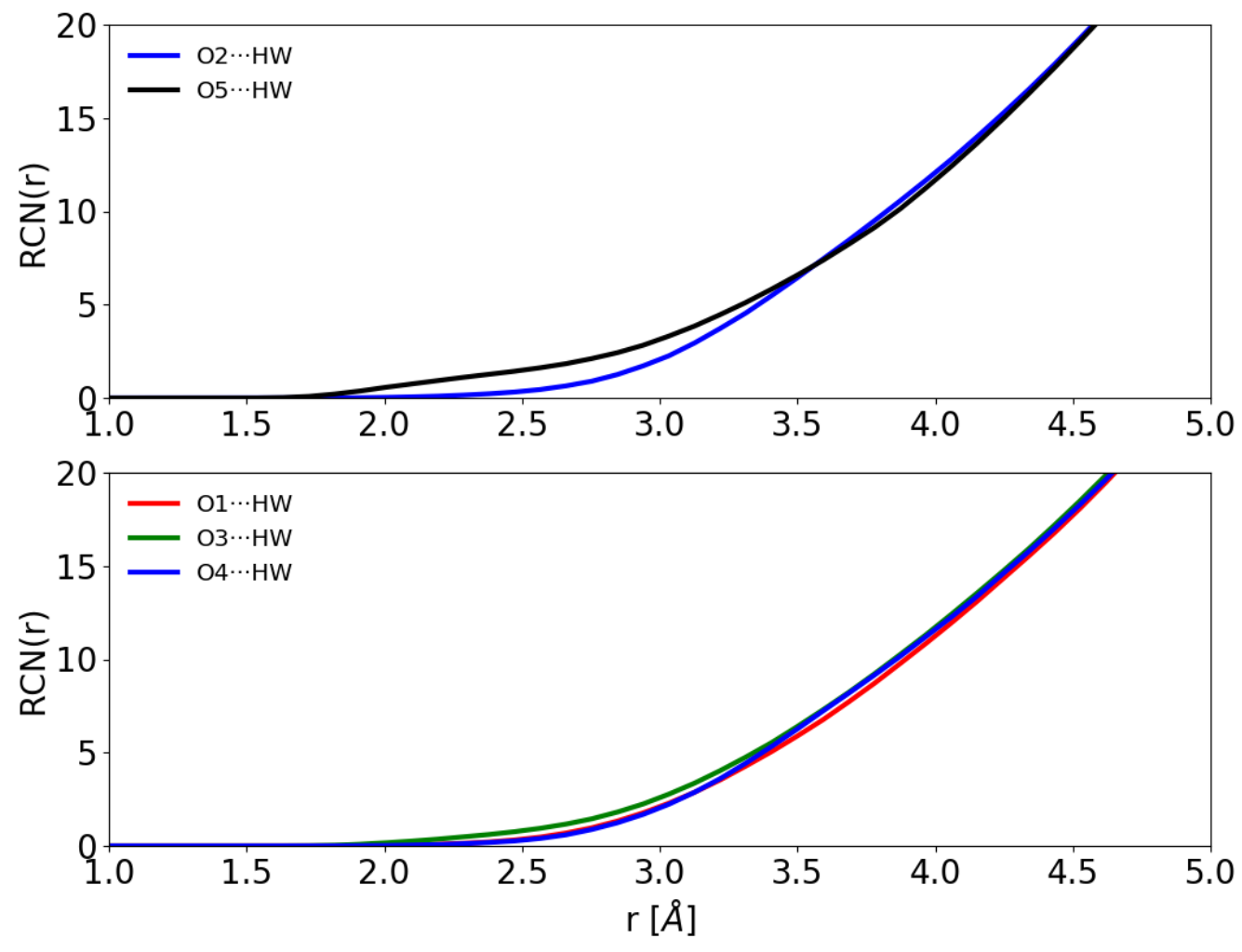}
\caption{GA radial distribution functions between GA oxygen atoms and water hydrogen atoms (HW). See Fig. \ref{fig:ga-struc} for atom labeling.}
\label{fig:rdf-O}
\end{figure}
\begin{figure}[htbp!]
\centering
\includegraphics[width=.48\textwidth]{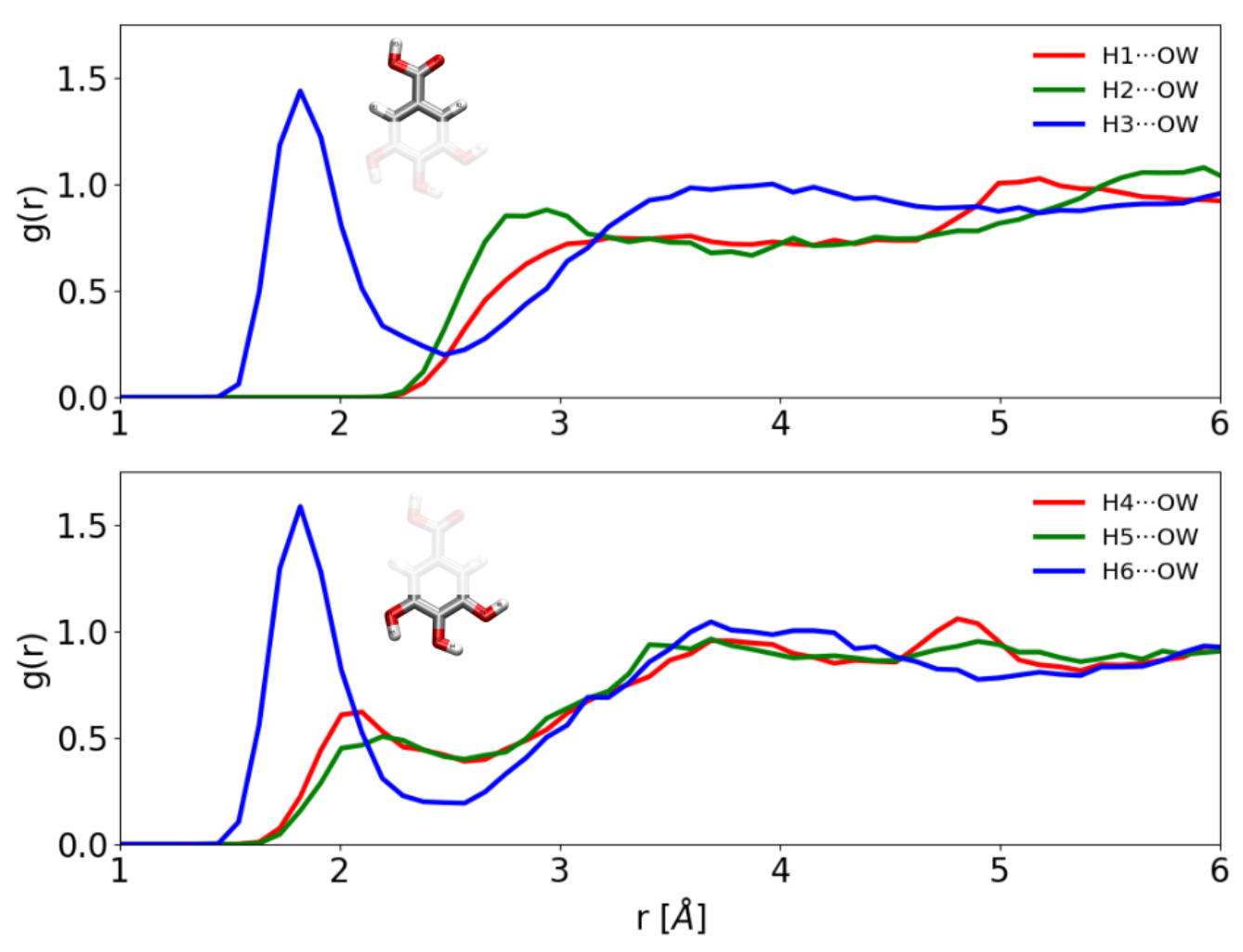}
\includegraphics[width=.48\textwidth]{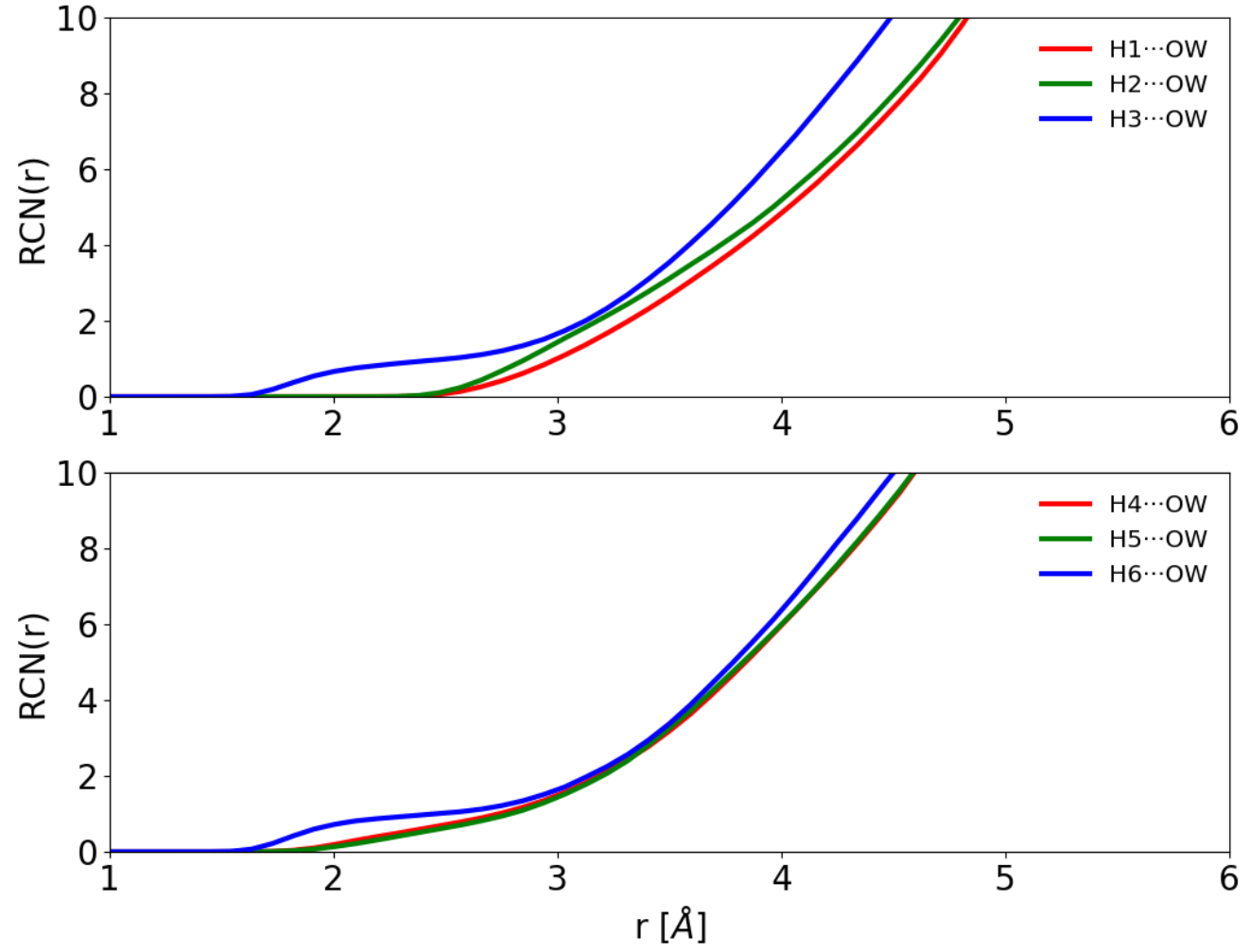}
\caption{GA radial distribution functions between GA hydrogen atoms and water oxygen atoms (OW). See Fig. \ref{fig:ga-struc} for atom labeling.}
\label{fig:rdf-H}
\end{figure}


before discussing GA IR spectra (see Fig. \ref{fig:ga-struc} for atom labelling), in this section the  MD trajectory is examined in terms of both radial distribution functions (rdfs) and running coordination numbers (RCNs). In particular, in order to obtain a description of the solvent local structure and to analyze hydrogen bonding patterns between GA and water molecules, intermolecular H(GA)$\cdots$OW and O(GA)$\cdots$HW rdfs and the corresponding running coordination numbers were calculated and are reported in Figs. \ref{fig:rdf-H} and \ref{fig:rdf-O} respectively. The coordination number of a specific site is defined by combining the distance at which the first/second minimum of rdf occurs with the corresponding distance in the running coordination number. 

The oxygen atoms of the three hydroxyl groups, i.e. O1, O3 and O4 (see Fig. \ref{fig:ga-struc}), present a radial distribution function with a similar shape, consisting of a broad peak at about 3.4 \AA, which is associated to the second solvation shell. The coordination number of these sites is equal to 16.7, 12.8 and 12.2, respectively. Remarkably, O3 rdf shows a peak at about 2.0 \AA, due to the fact that this is the only one among the hydroxyl groups that can form intermolecular HB with water molecules. O1 and O4 are involved in an intramolecular HB. Moving to the acid group, O5 rdf presents a sharp peak related to the first solvation shell, with a maximum at 2.0 \AA~ and a coordination number corresponding to 1.6; the plot for O2 is very similar to what has already been discussed in case of O1 and O4. %The rdf of O3 presents a shape that is somewhere in between the two previously commented shapes which can be ascribed to both geometry and charge distribution. Indeed the charge of the O3 atom is more negative, about 0.03, with respect to the O2 one, moreover the distance between this two atoms and the nearest hydrogen is also different (O2---H1: 2.4 \AA, O3---H1: 2.5 \AA). This two apparently negligible factors may have influenced the entire molecular dynamics. 

The specific arrangement of the hydroxyl groups in the meta and para positions (with respect to the acid) of GA induces a weak but detectable anisotropy on the rdfs of both hydrogen and oxygen atoms. In fact, both H4 and H5 rdfs present a similar shape, with a first peak at about 2.1 and 2.2 \AA, respectively. They correspond to coordination numbers of 0.8 and 0.7, respectively. A second peak related to the second solvation shell is placed at 3.68 \AA~, which corresponds to coordination numbers of 7.5 and 6.3, respectively. The H3 and H6, instead, show a pronounced first narrow peak at about the same distance (1.73 \AA), with a coordination number of 1.0 and a second peak around 3.97 \AA~ with coordination numbers of 15.2 and 13.8, respectively. Again, the differences between H3 and H4/H5 are due to the fact that H4 and H5 are involved in the intramolecular HB, whereas H6 is free to form intermolecular HB with water molecules. %To conclude the analysis of the A second solvation shell peak is clearly visible from the H2 rdf, but not for its symmetric H1 whose rdf remains approximatively flat until 5 \AA. 
Remarkably, the results here discussed are similar to the findings previously reported byone of the present authors.\cite{cappelli2005environmental}

\subsubsection{IR spectrum of GA in Aqueous Solution}

\begin{figure}[htbp!]
\centering
\includegraphics[width=.5\textwidth]{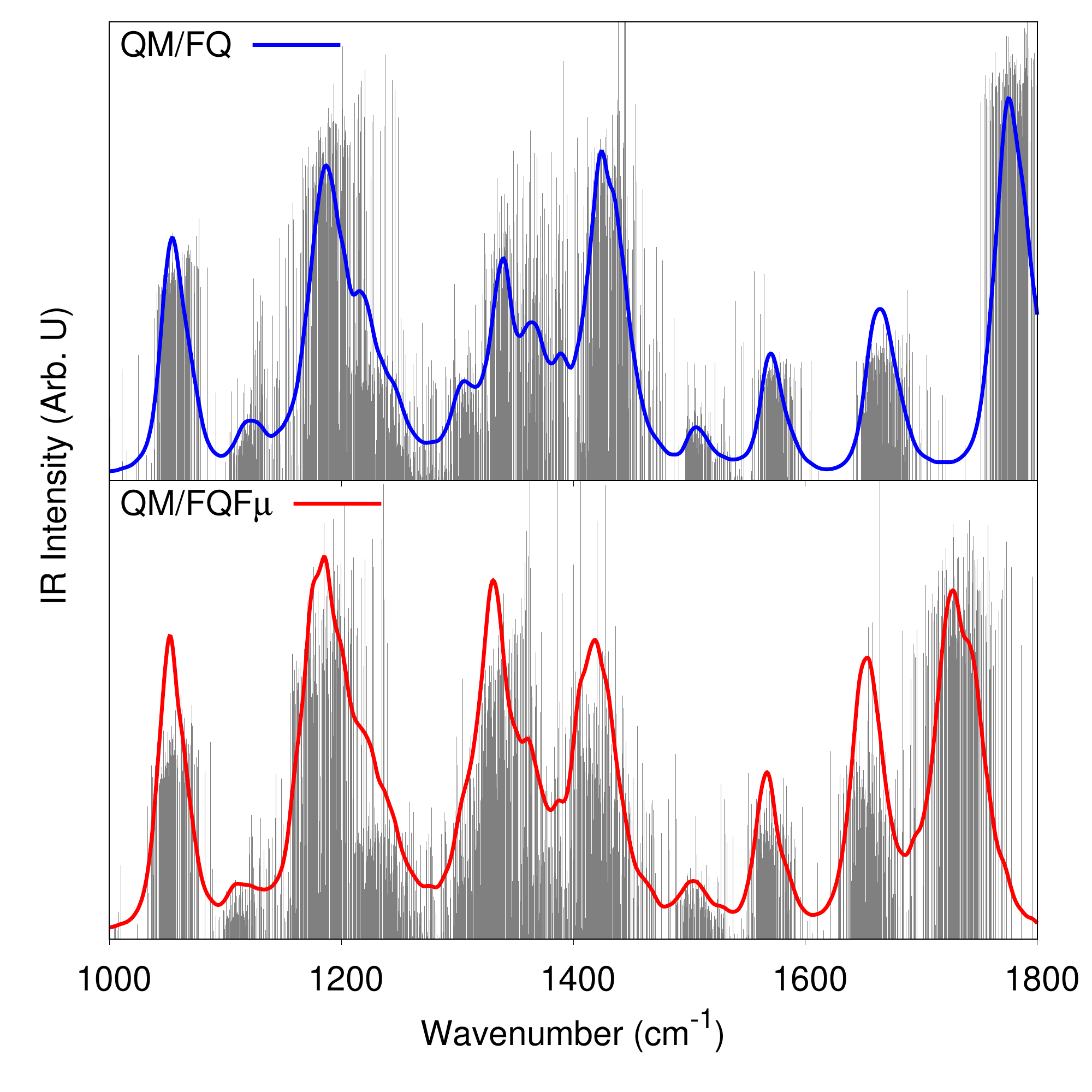}
\caption{Calculated QM/FQ (top) and QM/FQF$\mu$ (bottom) stick and convoluted IR spectra of GA in aqueous solution (FWHM = 4cm$^{-1}$.}
\label{fig:ga-stick}
\end{figure}

QM/FQ and QM/FQF$\mu$ IR spectra of GA were obtained by averaging over 200 snapshots extracted from the MD run. Similarly to MOXY and GLY, we checked that such a number of snapshots  produce converged spectra. The raw data extracted from the calculations, i.e. stick spectra are reported in Fig. \ref{fig:ga-stick} for the region 1000-1800 cm$^{-1}$  (i.e. the region of interest for the experimental investigation, vide infra); lorentzian convolution (FWHM = 4 cm$^{-1}$) is also depicted. Both QM/FQ and QM/FQF$\mu$ stick spectra show a large spreading in intensities and frequencies. Such a feature is reported for most computed bands, in particular in the region between 1100-1400 cm$^{-1}$, in which single bands are not easily detectable (see for comparison MOXY and GLY raw spectra in Figs. \ref{fig:moxy-stick} and \ref{fig:gly-stick}).  
%tommaso: la dinamica e' fissa come fatta da voi nell'articolo! QUI!!!!!!CHIARA: COMMENTO SULLA FLESSIBILITA' DEL GALLICO?????!!!!!!

The comparison between QM/PCM, QM/FQ, QM/FQF$\mu$ and the experimental IR spectrum \cite{lendl2002jacs} %tommaso: corretta, ringraziate dei tizi per l'IR QUI!!!! LA REF E' INESATTA 
shows that the latter is dominated by a three-band broad structure between 1200 and 1500 cm$^{-1}$, which probably involve more than one normal mode. 
A well-separated peak is present at 1000 cm$^{-1}$ and it is associated to the C-OH bending (see Fig. S3 in the SI for a graphical depiction of the normal modes). Moreover, three small bands of the same intensity are reported in the region 1500-1800 cm$^{-1}$, which are mainly due to composite C-OH bending modes of the hydroxyl groups and the acidic C=O stretching.

\begin{figure}[htbp!]
\centering
\includegraphics[width=.5\textwidth]{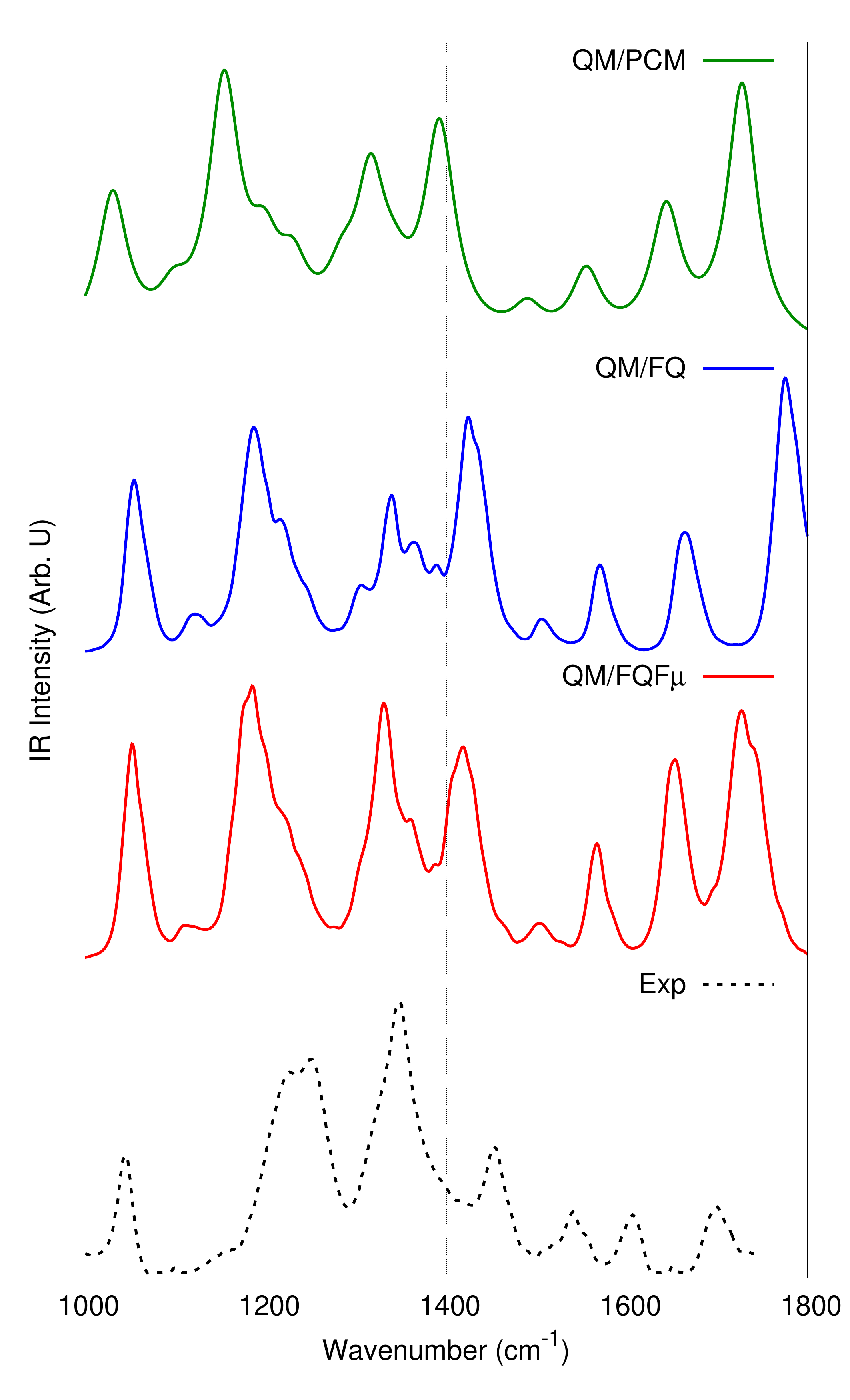}
\caption{Calculated QM/PCM (green), QM/FQ (blue) and QM/FQF$\mu$ (red) IR spectra of GA in aqueous solution (QM/PCM FWHM = 10 cm$^{-1}$; QM/MM FWHM = 4cm$^{-1}$). The experimental spectrum (dashed black line) is reproduced from Ref. \citenum{cappelli2005environmental}.}
\label{fig:ga-ir}
\end{figure}

%QUI!!!!!!!!!!!!!!CONFRONTO CON CLUSTER???????? tommaso: lo metto subito dopo dove non sei d'accordo!
%CONFRONTO CON NON-POLAR

The QM/PCM spectrum is dominated by two peaks, placed at about 1170 cm$^{-1}$ and 1750 cm$^{-1}$, respectively. Such peaks are related to a composite C-OH bending and to the C=O stretching. A similar spectrum is predicted by adopting the atomistic QM/FQ approach, in which the most intense peak is predicted in case of the C=O stretching at about 1750 cm$^{-1}$, whereas the peaks at about 1200 and 1400 cm$^{-1}$ have almost the same intensity. It is worth noticing that in QM/FQ spectra most bands present an inhomogenous broadening that is related once again to the dynamical picture given by the sampling of the phase-space through MD. In addition, similarly to MOXY and GLY, most of the computed QM/FQ bands are blueshifted with respect to their QM/PCM counterparts, thus reflecting the stronger solute-solvent interaction. 

Most QM/FQF$\mu$ bands are blueshifted with respect to QM/PCM, whereas they are redshifted with respect to QM/FQ values, thus highlighting the different electrostatic description given by the two explicit approaches. Remarkably, similarly to GLY in aqueous solution, vibrational frequencies are not completely in agreement with the experimental ones, that probably due to the lack of anharmonicity effects and the use of DFT. Moreover, inhomogenous band broadening is correclty repoduced by both atomistic approaches, thus resulting in a very good agreement with the experiments, in particular for the experimentally most intense band at about 1350 cm$^{-1}$. 

\begin{figure}[htbp!]
\centering
\includegraphics[width=.5\textwidth]{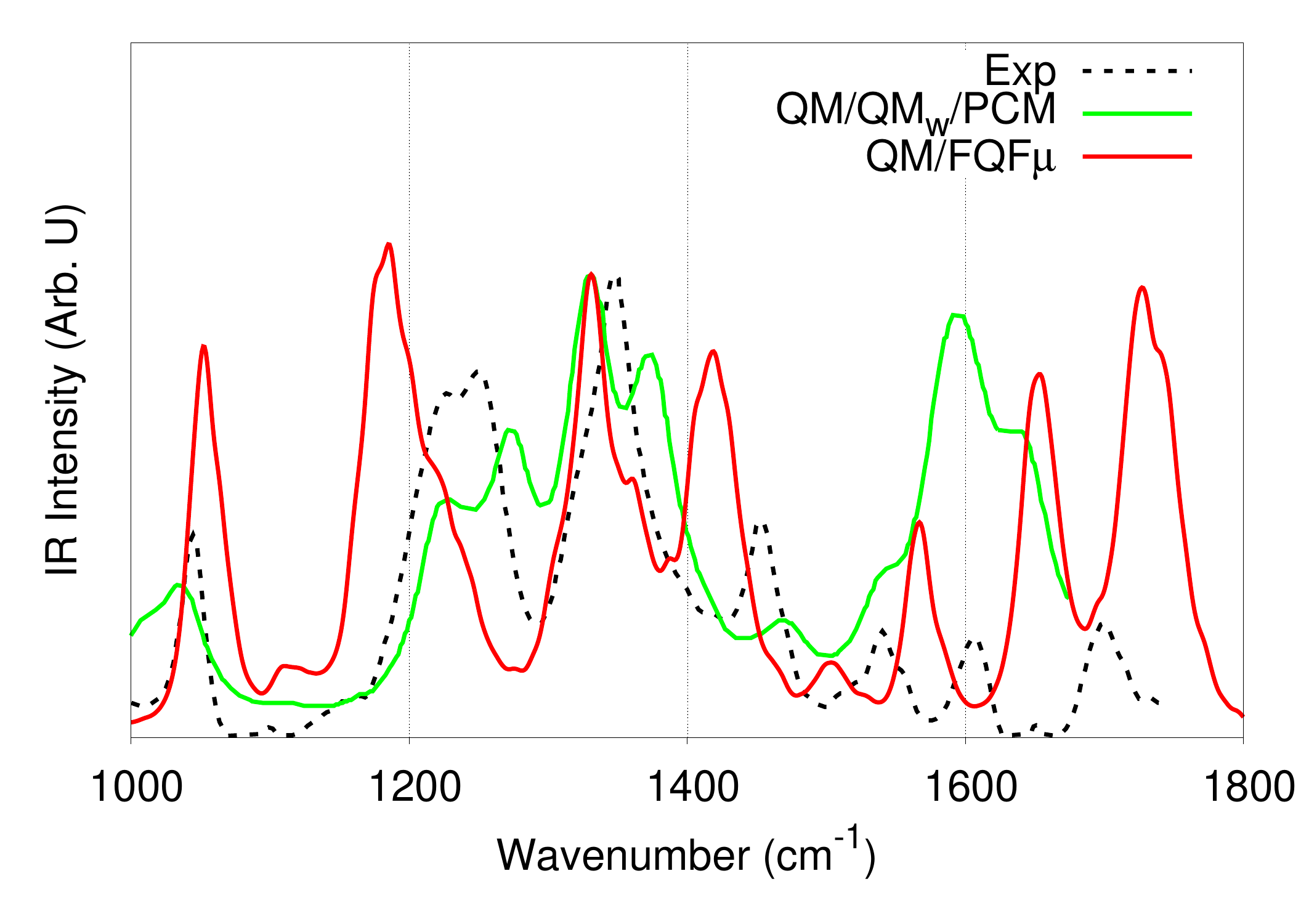}
\caption{Calculated QM/FQF$\mu$ (red) and QM/QM$_\text{w}$/PCM (green) reproduced form Ref. \citenum{cappelli2005environmental}. The experimental spectrum (dashed black line) is reproduced from Ref. \citenum{cappelli2005environmental}.}
\label{fig:ga-sovra}
\end{figure}

By further deepening the analysis of computed spectra, we see that QM/FQF$\mu$ IR spectrum is dominated by three bands at about 1200, 1350 and 1700 cm$^{-1}$, which are predicted almost with the same intensity. This is a big improvement with respect to both QM/PCM and QM/FQ approaches, because the most intense band in the experimental spectrum is correctly predicted only by QM/FQF$\mu$. It is worth noticing that a correct reproduction of the intensity of this peak was achieved by some of the present authors by resorting to a supermolecule approach (here called QM/QM$_\text{w}$/PCM), i.e. by including 8 QM water molecules in the definition of the QM solute in QM/PCM calculations (see Fig. \ref{fig:ga-sovra}). Such an observation indeed indicates that an explicit solvation approach is needed to recover the experimental features in this region. In addition, due to the fact that QM/FQF$\mu$ appropriately reproduces the most intense band of the experimental spectrum, in a similar way as the supermolecule approach, we can conclude that the electrostatic description of the HB interaction is the most relevant solute-solvent contribution, and that the electrostatic description given by QM/FQF$\mu$ in this case overcomes that modeled by the QM/FQ. The largest discrepancies between QM/FQF$\mu$ and QM/QM$_\text{w}$/PCM are reported for normal mode frequencies, which are better reproduced by the supramolecule approach, in particular in the region between 1200 and 1400 cm$^{-1}$. Such an improvement can be related to the fact that the supermolecule accounts for non-electrostatic interactions (in particular Pauli repulsion) which can therefore play a relevant role in the determination of vibrational frequencies.  
%tommaso: aggiungo lo spettro del 2005!!! e modificato la frase subito sopra! QUI!!!!!!!!!! CHIARA NON SONO D'ACCORDO.... SECONDO ME E' H BOND QUELLA PIà IMPORTANTE.... NELLA SUA COMPONENTE ELETTROSTATICA, MA H BOND.... NON MEDIA....
%QUI!!!!!!! IN CASO MODIFICARE SE METTIAMO LO SPETTRO DEL LAVORO DEL 2005

To conclude the discussion on GA, it is worth noticing that all the considered computational approaches predict very large intensities for the three modes in the 1500-1800 cm$^{-1}$ region, even the supramolecule QM/QM$_\text{w}$/PCM. This is probably due to both the lack of vibrational anharmonicity, which has been reported to affect not only frequencies but also intensities.\cite{cappelli2011towards,bloino2012second,bloino2015anharmonic} 

\section{Summary and Conclusions}

In this work, the fully polarizable QM/FQF$\mu$ approach, recently developed by some of the present authors, has been extended to the evaluation of IR spectra, though the development of analytical energy first and second derivatives. In QM/FQF$\mu$ both a charge and a dipole, which can vary as a response to the external electric field and potential, are placed on each atom of the MM portion. Such a model can be viewed as a refinement of the QM/FQ approach, in which only fluctuating charge are considered. 

Our approach has been tested against the reproduction of IR spectra of three systems in aqueous solution, namely methyloxirane, glycidol and gallic acid. The selected molecules can interact with water through  strong solute-solvent interactions; this is reflected in the computed IR spectra by the fact that the atomistic QM/FQ and QM/FQF$\mu$ generally over-perform the implicit QM/PCM. In particular, and as expected, the bands which are mostly affected by the atomistic description of the environment, are those involving the polar moieties of the investigated molecular systems. 

In case of both methyloxirane and glycidol in aqueous solution, QM/FQF$\mu$ predicts similar spectra with respect to QM/FQ, whereas for gallic Acid the inclusion of anisotropic terms in the MM modeling, i.e. the inclusion of fluctuating dipoles, permits to obtain a better agreement with the experimental data. In the latter case, we also noticed that some bands can probably be affected by anharmonicity\cite{yagi2019anharmonic} and non-electrostatic solute-solvent interactions. The extension of QM/FQF$\mu$ so to include anharmonicity and non-electrostatic interactions, for instance by extending the method already developed by some of the present authors,\cite{giovannini2017disrep,giovannini2019eprdisrep,curutchet2018density} might be beneficial and will be the subject of future communications. 
Moreover, it is worth pointing out that the development and implementation of analytical first energy derivatives, i.e. energy gradients, is not only the basic ingredient for computing vibrational spectra, but it also allows for a further extension of the model to QM/MM MD, as already reported for other kinds of polarizable QM/MM approaches.\cite{thompson1996qm,loco2017hybrid,kratz2016lichem,dziedzic2016tinktep}

\newpage

\section{Appendix}

\subsection{FQF$\mu$ Energy First Derivatives with respect to MM coordinates}

The derivative of the energy with respect to the position of an MM atom, which we will denote with the superscript $\xi$, can be obtained by using the chain rule. Only explicit contributions arise, as the overlap matrix does not depend on  MM atom positions. In fact,

\begin{equation}
\mathcal{E}^{\xi} = \frac{1}{2}\mathbf{q}^{\dagger}\mathbf{T}^{\xi}_{qq}\mathbf{q} + \frac{1}{2}\bm{\mu}^{\dagger}\mathbf{T}^{\xi}_{\mu\mu}\bm{\mu} + \mathbf{q}^{\dagger}\mathbf{T}^{\xi}_{q\mu}\bm{\mu} + \mathbf{q}^{\dagger}\mathbf{V}^{\xi}(\mathbf{P}) - \bm{\mu}^{\dagger}\mathbf{E}^{\xi}(\mathbf{P})
\label{eq:1derMM}
\end{equation}

where the derivative of the QM/MM interaction potential is equal to the electric field produced by the QM density acting on the charges, whereas the derivative of the QM/MM interaction field is the electric field gradient acting on the dipoles.

The derivatives of the interactions kernels $\mathbf{T}^{\xi}_{qq}$, $\mathbf{T}^{\xi}_{q\mu}$ and $\mathbf{T}^{\xi}_{\mu\mu}$ can be obtained by differentiating Eqs. \ref{eq:kernelqq}, \ref{eq:Tqmu} and \ref{eq:Tmumu}:

\begin{align}
\text{T}^{\xi}_{qq, ij} & = -\text{T}^{q\mu}_{ij} \\
\text{T}^{\xi}_{q\mu, ij} & = -\text{T}^{q\mu}_{ij} \\
\text{T}^{\xi}_{\mu\mu, ij} & = \left \{ \frac{3}{\abs{\mathbf{r}_{ij}}^5}\left [ \erf\left(\frac{\abs{\mathbf{r}_{ij}}}{R}\right) -\frac{2}{\sqrt{\pi}}\frac{\abs{\mathbf{r}_{ij}}}{R}\;\exp\left(-\frac{\abs{\mathbf{r}_{ij}}^2}{R^2}\right)\right ] + \right. \nonumber \\
& \left. -\frac{4}{\sqrt{\pi}R^3}\frac{1}{\abs{\mathbf{r}_{ij}}^2}\;\exp\left(-  \frac{\abs{\mathbf{r}_{ij}}^2}{R^2}\right) \right \}\mathbf{K}^{(x_i)}+ \nonumber \\
& +\frac{\text{K}_{xx}}{\abs{\mathbf{r}_{ij}}}\Bigg \{ \frac{3\abs{\mathbf{r}_{ij}}^2 \mathbf{I}-15\mathbf{K}}{\abs{\mathbf{r}_{ij}}^6}\left [ \erf\left(\frac{\abs{\mathbf{r}_{ij}}}{R}\right) -\frac{2}{\sqrt{\pi}}\frac{\abs{\mathbf{r}_{ij}}}{R}\;\exp\left(-\frac{\abs{\mathbf{r}_{ij}}^2}{R^2}\right)\right ] + \nonumber \\
& +\frac{3\mathbf{K}-\abs{\mathbf{r}_{ij}}^2 \mathbf{I}}{\abs{\mathbf{r}_{ij}}^5}\frac{2}{\sqrt{\pi}}\frac{1}{R}\exp\left(-\frac{\abs{\mathbf{r}_{ij}}^2}{R^2}\right)\left [ \frac{2\abs{\mathbf{r}_{ij}}^2}{R^2}-1-\frac{2\abs{\mathbf{r}_{ij}}}{R} \right ]+ \nonumber\\
& +\frac{8}{\sqrt{\pi}R^5}\frac{\mathbf{K}}{\abs{\mathbf{r}_{ij}}^3}\;\exp\left(-\frac{\abs{\mathbf{r}_{ij}}^2}{R^2}\right) \left ( R^2+\abs{\mathbf{r}_{ij}}^2 \right )\Bigg \}
\label{eq:tmumu-der}
\end{align}

where $\mathbf{K} = \gr{r}_{i,j}\otimes\gr{r}_{i,j}$ and $\mathbf{K}^{(x_i)}$ is the derivative of $\mathbf{K}$ with respect to the $x$ component of the $i$-th element. For the sake of clarity, in Eq. \ref{eq:tmumu-der} $R_{\mu_i-\mu_j}$ was substituted by $R$.

\subsection{FQF$\mu$ Energy Second Derivatives with respect to MM coordinates}

For the sake of completeness, in this appendix the formulas for the full Hessian matrix, i.e. including also the QM/MM and MM/MM blocks, are given. Derivatives with respect to MM
atom coordinates will be denoted by the superscripts $\xi,\eta$. The QM-MM block of the Hessian can be obtained by differentiating once the forces on the MM portion with respect to the position of a QM nucleus:

\begin{equation}
\mathcal{E}^{(x\xi)} (\mathbf{P},\mathbf{q},\bm{\mu},\bm{\lambda}) = \parz{\mathcal{E}^{\xi}}{x} + \parz{\mathcal{E}^{\xi}}{\mathbf{P}}\parz{\mathbf{P}}{x} + \parz{\mathcal{E}^{\xi}}{\mathbf{q}}\parz{\mathbf{q}}{x} + \parz{\mathcal{E}^{\xi}}{\bm{\mu}}\parz{\bm{\mu}}{x} +  \parz{\mathcal{E}^{\xi}}{\bm{\lambda}}\parz{\bm{\lambda}}{x}
\end{equation}

where the last term vanishes. Substituting $\mathcal{E}^{\xi}$ with Eq \ref{eq:1derMM}:

\begin{align}
\mathcal{E}^{(x\xi)} &= \mathbf{q}^{\dagger}\mathbf{V}^{x\xi}(\mathbf{P}) + \mathbf{q}^{\dagger}\mathbf{V}^{\xi}(\mathbf{P}^x) - \bm{\mu}^{\dagger}\mathbf{E}^{x\xi}(\mathbf{P} - \bm{\mu}^{\dagger}\mathbf{E}^{\xi}(\mathbf{P}^x) + \nonumber \\
 &+ \left(\mathbf{T}^{\xi}_{qq}\mathbf{q}+\mathbf{T}^{\xi}_{q\mu}\bm{\mu} + \mathbf{V}^{\xi}(\mathbf{P})\right)^{\dagger}\mathbf{q}^x + \left(\mathbf{T}^{\xi}_{\mu\mu}\bm{\mu}+\mathbf{q}^{\dagger}\mathbf{T}^{\xi}_{q\mu} - \mathbf{E}^{\xi}(\mathbf{P})\right)^{\dagger}\bm{\mu}^x
\end{align}

The derivatives of the density matrix and of the FQs can be obtained by solving the CPHF equations described in Section 2.2.2: therefore, there is no need to enlarge the CPHF system of equations to calculate the derivatives of the density matrix with respect to the positions of the MM atoms. This is, however, unavoidable when the MM-MM block of the Hessian is to be calculated. By differentiating Eq. \ref{eq:1derMM} with respect to the position of MM atoms:

\begin{equation}
\mathcal{E}^{(\xi\eta)} (\mathbf{P},\mathbf{q},\bm{\mu},\bm{\lambda}) = \parz{\mathcal{E}^{\xi}}{\eta} + \parz{\mathcal{E}^{\xi}}{\mathbf{P}}\parz{\mathbf{P}}{\eta} + \parz{\mathcal{E}^{\xi}}{\mathbf{q}}\parz{\mathbf{q}}{\eta} + \parz{\mathcal{E}^{\xi}}{\bm{\mu}}\parz{\bm{\mu}}{\eta} + \parz{\mathcal{E}^{\xi}}{\bm{\lambda}}\parz{\bm{\lambda}}{\eta}
\end{equation}

The last term vanishes once again; to calculate the derivatives of the charges and the density, a new set of CPHF equations needs to be solved. Differentiation of the Liouville equation (as the overlap does not depend on the positions of the MM atoms, we can work in the MO basis) and projection onto the o-v block gives:

\begin{equation}
\mathbf{FP}^{\xi}_{ov} - \mathbf{P}^{\xi}_{ov}\mathbf{F} = \mathbf{F}^{\xi}_{ov}
\end{equation}

The Fock matrix derivatives have no contributions arising from the one- and two-electron matrices, but only from the density and FQF$\mu$ derivatives:

\begin{equation}
F^{\xi}_{ia} = G_{ia}(\mathbf{P}^{\xi}) + \mathbf{q}^{\xi\dagger}\mathbf{V}_{ia} + \mathbf{q}^{\dagger}\mathbf{V}^{\dagger}_{ia} - \bm{\mu}^{\xi\dagger}\mathbf{E}_{ia} - \bm{\mu}^{\dagger}\mathbf{E}^{\dagger}_{ia} 
\end{equation}

By differentiating the FQF$\mu$ equations we obtain:

\begin{equation}
\mathbf{D}^{\xi} \mathbf{L} + \mathbf{D}\mathbf{L}^{\xi} = -\mathbf{R}^{\xi}(\mathbf{P}) - \mathbf{R}(\mathbf{P}^{\xi})
\end{equation}

By putting everything together, a Casida-like system of equations is obtained, where the matrices are defined in Eq. 22 and the right-hand side reads:

\begin{equation}
\mathbf{Q}^{\eta}_{ia} = -\mathbf{L}^{\dagger}\mathbf{R}^{\xi}_{ia} + \mathbf{R}^{\dagger}_{ia}\mathbf{D}^{-1}(\mathbf{D}^{\xi}\mathbf{L} + \mathbf{R}^{\xi}(\mathbf{P}))
\end{equation}

\section{Supporting Information}

Graphical depiction of normal modes of MOXY, GLY and GA in aqueous solution. %tommaso: non dovrebbe! QUI!!!!!!!!! CONTROLLA SE C'E' ALTRO!!!!!!!!!!!!!!!!!!!!!!!!

\section{Acknowledgments}

We are thankful for the computer resources provided by the high performance computer facilities of the SMART Laboratory (http://smart.sns.it/). CC gratefully acknowledges the support of H2020-MSCA-ITN-2017 European Training Network “Computational Spectroscopy In Natural sciences and Engineering” (COSINE), grant number 765739. 
\newpage

{
\small
\bibliography{biblio}

\begin{mcitethebibliography}{83}
\bibitem[Lin and Truhlar(2007)Lin, and Truhlar]{lin2007qm}
Lin,~H.; Truhlar,~D.~G. {QM/MM}: what have we learned, where are we, and where
  do we go from here? \emph{Theor. Chem. Acc.} \textbf{2007}, \emph{117},
  185--199\relax
\mciteBstWouldAddEndPuncttrue
\mciteSetBstMidEndSepPunct{\mcitedefaultmidpunct}
{\mcitedefaultendpunct}{\mcitedefaultseppunct}\relax
\EndOfBibitem
\end{mcitethebibliography}
}

\end{document}